\documentclass[11pt]{article}

\usepackage[utf8]{inputenc}
\usepackage[T1]{fontenc}
\usepackage{lmodern}
\usepackage{geometry}
\usepackage{microtype}
\usepackage{graphicx}
\usepackage{multirow}
\usepackage{amsmath,amssymb,amsfonts}
\usepackage{amsthm}
\usepackage{mathrsfs}
\usepackage[title]{appendix}
\usepackage{xcolor}
\usepackage{textcomp}
\usepackage{manyfoot}
\usepackage{booktabs}
\usepackage{algorithm}
\usepackage{algorithmicx}
\usepackage{algpseudocode}
\usepackage{listings}
\usepackage{float}
\usepackage{placeins}
\usepackage{authblk}
\usepackage[numbers,sort&compress]{natbib}
\usepackage{hyperref}
\hypersetup{colorlinks=true,linkcolor=darkblue,citecolor=darkblue,urlcolor=darkblue}

\definecolor{rev}{rgb}{0,0,120}
\definecolor{darkblue}{rgb}{0.0,0.0,0.55}
\newcommand*{\BT}{\textcolor{black}}

\theoremstyle{plain}

\theoremstyle{remark}

\theoremstyle{definition}

\begin{document}

\title{Accelerating Bayesian inverse design in computational fluid dynamics using neural operators}

\author[1]{Bipin Tiwari\thanks{Email: \texttt{btiwari1@vols.utk.edu}}}
\author[1]{Omer San\thanks{Corresponding author. Email: \texttt{osan@utk.edu}}}
\affil[1]{Department of Mechanical and Aerospace Engineering, University of Tennessee, Knoxville, TN 37996, USA}
\date{}

\maketitle

\begin{abstract}

\BT{Bayesian inverse design provides a principled framework for inferring aerodynamic geometries from sparse flow observations while quantifying uncertainty. However, its practical use in computational fluid dynamics (CFD) is severely limited by the cost of repeated high-fidelity simulations required for gradient-based Markov chain Monte Carlo (MCMC) sampling.} While surrogate models are commonly proposed to reduce this cost, their effect on posterior geometry and uncertainty, especially for shock-dominated flows, remains poorly understood. \BT{In this work, we demonstrate that neural operator surrogates can be embedded directly within the MCMC inference loop while preserving posterior structure.} Using a fully Bayesian inverse formulation of quasi-one-dimensional nozzle flow, we demonstrate that geometry parameterization plays a decisive role in identifiability and posterior conditioning, with cubic B-splines yielding stable and physically meaningful uncertainty estimates. Building on this formulation, a Deep Operator Network trained on CFD-generated data is substituted for the CFD solver within a No-U-Turn Sampler, while keeping the likelihood model, priors, and sampling configuration unchanged. Across sparse to fully observed regimes, surrogate-based inference reproduces the posterior geometry and uncertainty trends of the CFD reference. As a result of surrogate integration, total inference time is reduced to under one second, corresponding to a speedup exceeding three orders of magnitude. In addition, a direct inverse neural operator is examined as a deterministic alternative for inverse design, enabling single-shot geometry reconstruction without posterior sampling. These results demonstrate that neural operator--accelerated Bayesian inference enables practical, uncertainty-aware inverse design workflows for aerodynamic applications.
\end{abstract}

\noindent\textbf{Keywords:} Inverse Design, Bayesian Inference, Scientific Machine Learning, Computational Fluid Dynamics, Uncertainty Quantification

\section{Introduction}
High-speed aerospace systems, including hypersonic vehicles, launch platforms, and atmospheric reentry capsules, operate in regimes where flow behavior is highly sensitive to geometric variations~\cite{boccaletto2010high,hadjadj2009nozzle, gamble2004nozzle, beans1992nozzle, goeing1990nozzle, hunter1998experimental}. Even small perturbations in nozzle contour can significantly influence compressible flow features such as shock formation, flow acceleration, and thrust efficiency~\cite{Anderson1995, korte2000inviscid}. This strong, nonlinear geometry--flow coupling renders inverse nozzle design inherently ill-posed, particularly when flow observations are sparse or contaminated by noise. As a result, nozzle design remains a central challenge in high-speed aerodynamics, particularly as modern aerospace systems demand higher performance, wider operating envelopes, and increased robustness under uncertainty~\cite{bertin2006critical,roy2011comprehensive}. 

High-fidelity computational fluid dynamics (CFD) has become the primary tool for analyzing and designing nozzle flows, as experimental access to extreme Mach number regimes is limited and costly~\cite{candler2019rate, knight2012assessment}. Modern solvers based on Reynolds-averaged or large-eddy Navier--Stokes formulations can resolve detailed flow features in convergent--divergent nozzles, including shock cells, flow separation, and turbulent structures~\cite{boles2009simulations,georgiadis2006navier,vos2002navier,hadjadj2005computation}. However, CFD-based design pipelines are computationally intensive, especially when used in iterative optimization or uncertainty analysis~\cite{skinner2018state,forrester2009recent}. This limitation has motivated the development of inverse design approaches, in which the nozzle geometries are inferred directly from prescribed flow characteristics such as the target Mach or pressure distributions~\cite{sivells1970aerodynamic, morris2012adjoint}. Most existing inverse design methods are deterministic, yielding a single reconstructed geometry and providing limited insight into non-uniqueness, sensitivity to noise, or uncertainty in the inferred design~\cite{schillings2011efficient}.

Bayesian inverse design addresses these limitations by reformulating geometry inference as a probabilistic problem, yielding posterior distributions that quantify uncertainty and accommodate non-uniqueness arising from sparse or noisy observations~\cite{kirsch2011introduction, kaipio2005statistical, stuart2010inverse}. For PDE-constrained inverse problems, gradient-based Markov Chain Monte Carlo (MCMC) methods, particularly Hamiltonian Monte Carlo (HMC) and its adaptive variants, are attractive due to their scalability in high-dimensional parameter spaces~\cite{brooks2011handbook,neal2011mcmc,hoffman2014no,peherstorfer2018survey,kaipio2005statistical}. Unlike traditional random-walk samplers that explore blindly, HMC adopts an analogy to physical dynamics: it introduces auxiliary momentum variables to evolve the state of the chain across a potential energy surface defined by the negative log-posterior~\cite{neal2011mcmc, brooks2011handbook}. To ensure robust performance without manual tuning, modern implementations utilize the No-U-Turn Sampler (NUTS), which adaptively sets the integration time to prevent the trajectory from doubling back on itself~\cite{hoffman2014no}. These methods exploit gradient information of the log-posterior to generate efficient proposals, making them especially attractive for PDE-constrained inverse problems. Despite their efficiency, gradient-based MCMC methods remain computationally infeasible when each likelihood evaluation requires a full CFD simulation. Thousands of forward solves are typically required to adequately explore the posterior, and the cumulative cost of solving the governing PDE dominates the inference process. As a result, the direct application of fully Bayesian methods to realistic aerodynamic inverse problems has remained largely out of reach. To mitigate this bottleneck, surrogate-augmented Bayesian inference strategies have been proposed, including reduced-order models and Gaussian process surrogates. While effective in low-dimensional or weakly nonlinear settings, these approaches often struggle to scale to strongly nonlinear PDEs with localized features such as shocks~\cite{christen2005markov,williams2006gaussian,marzouk2009dimensionality,benner2015survey,san2015stabilized,san2018extreme, tiwari2025epistemic}. This motivates surrogates that are both expressive and globally valid over parameter space, while remaining differentiable so that they can be used inside gradient-based samplers without compromising the posterior structure.

Neural operators enable this by learning mappings between function spaces (operators) instead of fitting solutions at individual parameter values. This yields fast, differentiable, and globally defined forward evaluations across a distribution of geometries, which aligns naturally with the needs of HMC/NUTS. Within scientific machine learning, both physics-informed neural networks (PINNs) and operator-learning methods have been introduced as alternatives to traditional solvers~\cite{raissi2019physics, cai2021physics, wang2021understanding}. Although both approaches generalize beyond their training data, PINNs typically operate at the level of individual solution fields, while operator-learning methods learn mappings across parameter space, making the latter more efficient for repeated forward evaluations in MCMC-based inference~\cite{Goswami2023, kovachki2023neural,lu2021learning}. Among operator-learning approaches, Deep Operator Networks (DeepONet) provide non-iterative forward evaluations and are fully differentiable, enabling direct integration within gradient-based MCMC samplers, enabling surrogate-accelerated Bayesian inference while preserving the underlying probabilistic formulation.~\cite{lu2021learning, wang2021learning}. DeepONets, along with more recent basis-to-basis and Laplace neural operator formulations, learn mappings between function spaces and have shown enhanced expressiveness and generalization capabilities for parametric PDE problems~\cite{ingebrand2025basis, cao2024laplace, goswami2024learning}.

\begin{figure}[H]
  \centering
  \includegraphics[width=\linewidth]{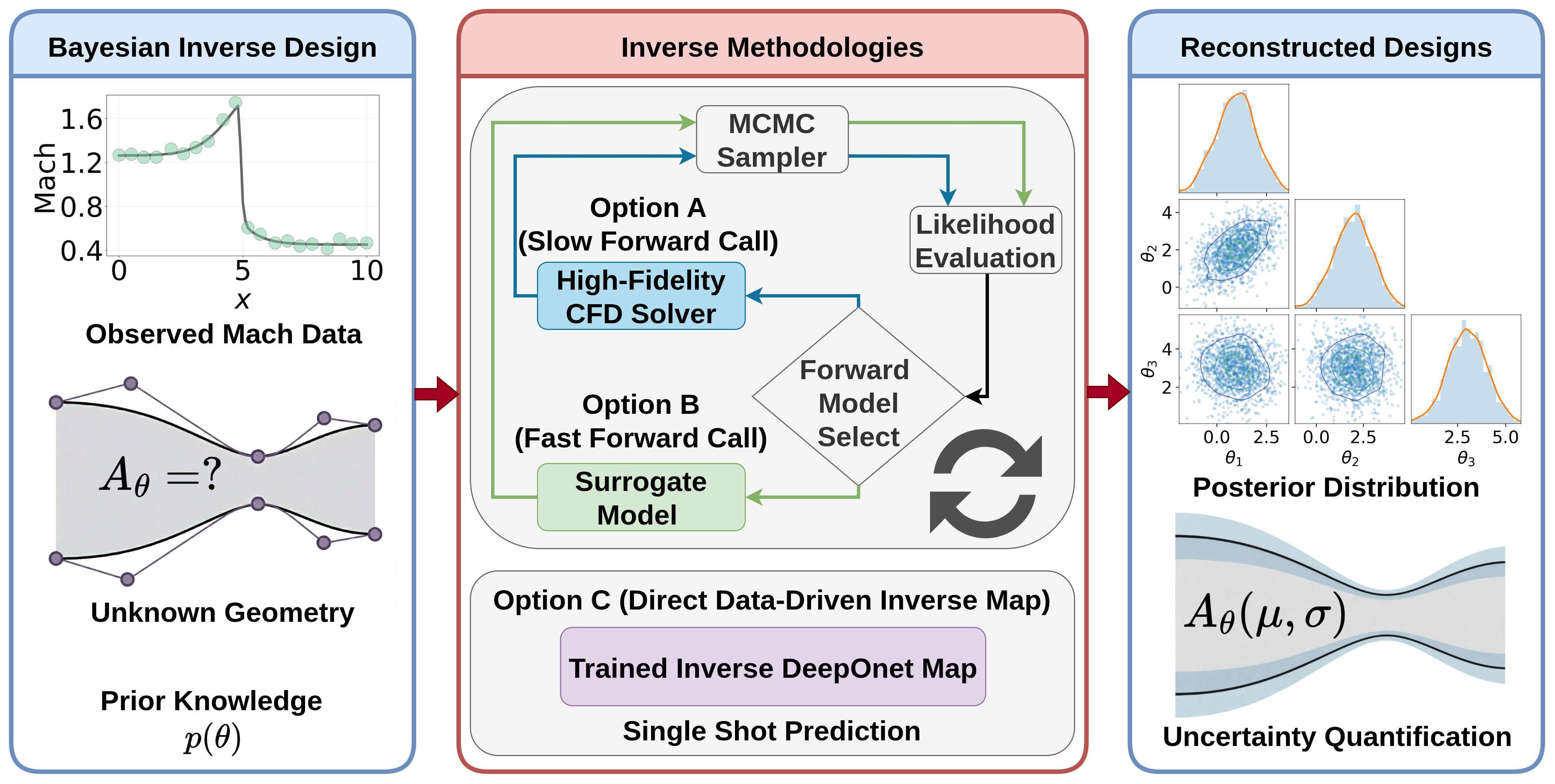}
  \caption{Overview of inverse nozzle design strategies compared in this work. Left: Bayesian inverse design infers an unknown geometry from noisy Mach observations under a prior. Middle: inverse methodologies include (A) CFD-based Bayesian inference using gradient-based MCMC, (B) surrogate-accelerated Bayesian inference in which a DeepONet replaces the CFD forward model inside the likelihood, and (C) a direct data-driven inverse DeepONet that predicts geometry parameters from observations via single-shot inference. Right: reconstructed designs and uncertainty quantification, including posterior distributions and credible envelopes (Approaches A--B) and reconstructed geometry from direct inversion (Approach C).}
  \label{fig:inverse_methods_overview}
\end{figure}

Beyond surrogate-accelerated Bayesian inference, an alternative strategy is to learn a direct data-driven inverse mapping from observed flow quantities to geometry parameters, enabling single-shot inversion after training. Such approaches trade posterior uncertainty quantification for extreme computational efficiency and are therefore useful as alternatives.
Fig.~\ref{fig:inverse_methods_overview} provides a conceptual overview of the inverse-design framework considered in this work and the methodology proposed within a broader spectrum of inverse approaches. The inverse problem begins with sparse and noisy flow observations, such as Mach number distributions, together with prior knowledge of the unknown nozzle geometry or its parameterization. These ingredients define a Bayesian inverse problem in which uncertainty in the data and modeling assumptions is propagated to uncertainty in the inferred design.

Leveraging these advantages of neural operator models, the primary objective of this study is to enable practical, uncertainty-aware inverse nozzle design through surrogate-accelerated Bayesian inference without compromising posterior fidelity. The contributions of this work are threefold. First, we establish a CFD-based Bayesian reference for inverse nozzle design using multiple geometry parameterizations, including a discrete representation, a polynomial model, and a cubic B-spline, and identify the cubic B-spline as providing the most stable posterior behavior and lowest reconstruction error. Second, we demonstrate that a DeepONet can replace the CFD solver within a gradient-based MCMC framework while preserving posterior mean structure and uncertainty trends across sparse to fully observed regimes. Third, we implement a direct inverse DeepONet as a deterministic alternative approach, highlighting the trade-offs between single-shot reconstruction and uncertainty quantification.

The remainder of this paper is organized as follows. 
Section~\ref{sec:methodology} describes the quasi--one-dimensional nozzle flow forward model, the geometry parameterization considered, the Bayesian formulation of the inverse problem,and the integration of a Deep Operator Network within a gradient-based MCMC sampling strategy as well as the direct inverse neural operator. 
Section~\ref{sec:results} presents numerical results, beginning with a CFD-based Bayesian reference study used to examine the effect of geometry parameterization, followed by a systematic comparison between CFD-based and DeepONet-accelerated Bayesian inference across sparse to fully observed measurement regimes. Reconstruction accuracy and computational efficiency are assessed using posterior statistics and error metrics, and a deterministic direct inverse neural operator is also examined as an alternative approach for inverse design. 
Finally, Section~\ref{sec:conclusions} summarizes the main findings, discusses limitations of the proposed framework, and outlines directions for future extensions to higher-dimensional, unsteady, and multi-fidelity inverse design problems.

\FloatBarrier

\section{Methodology}
\label{sec:methodology}

This section describes the computational framework used to solve the inverse nozzle design problem, in which the nozzle geometry is inferred from flow observations through a nonlinear PDE-constrained forward model. The goal is to enable uncertainty-aware inference while maintaining computational tractability. To this end, we consider three complementary inverse strategies that share the same physical model, observation setting, and geometry parameterization but differ in how forward and inverse mappings are evaluated.

The methodology comprises the definition of the forward flow operator, finite-dimensional parameterization of the nozzle geometry, Bayesian formulation of the inverse problem, and the construction of neural-operator surrogates for forward and inverse mappings. This unified framework allows for a controlled and systematic comparison between probabilistic Bayesian inference with the CFD-based forward solver and the neural operator-based forward solver. In addition, a single-shot deterministic direct inversion method is presented as an alternative.

\begin{figure}[H]
  \centering
  \includegraphics[width=0.95\textwidth]{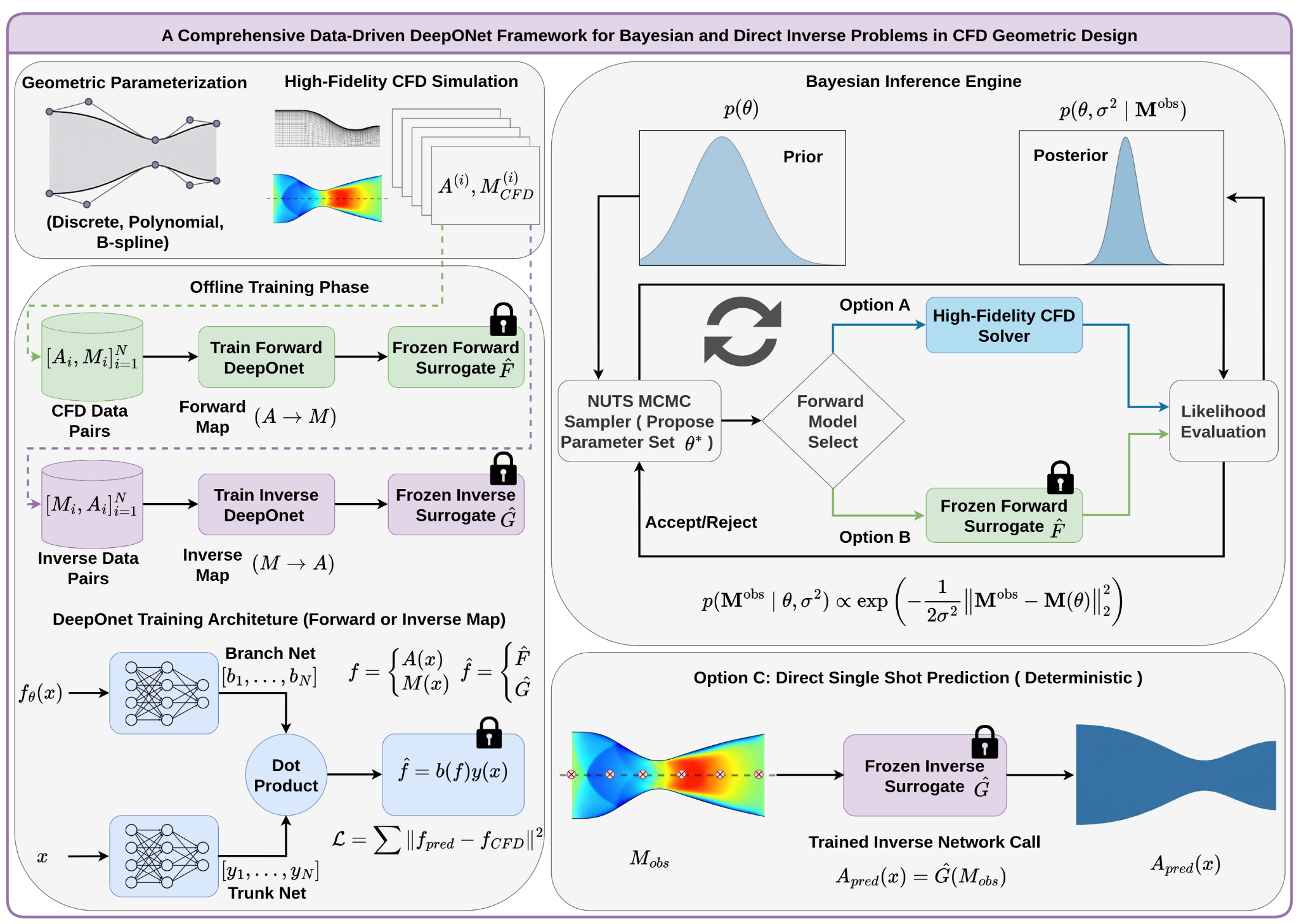}
  \caption{
  Comprehensive DeepONet-based inverse-design framework used in this study.
  Geometry parameterizations are used to generate paired CFD datasets for offline training.
  Forward and inverse DeepONets are trained and frozen prior to inference.
  The frozen forward surrogate is integrated within a Bayesian inference engine using a NUTS MCMC sampler, enabling either high-fidelity CFD-based likelihood evaluation (Option A) or surrogate-accelerated Bayesian inference (Option B).
  In parallel, a frozen inverse DeepONet provides direct single-shot geometry prediction from observed flow data (Option C).
  }
  \label{fig:methodology_overview}
\end{figure}

Fig.~\ref{fig:methodology_overview} provides a schematic overview of the entire workflow, highlighting the separation between offline training and online inference in the left panels. In right panels of Fig.~\ref{fig:methodology_overview}, three inverse-design approaches are illustrated in detail: (A) fully Bayesian inference using a high-fidelity CFD solver as the forward model; (B) surrogate-accelerated Bayesian inference in which the CFD solver is replaced by a neural operator surrogate; and (C) direct data-driven inverse prediction, where a trained neural operator maps flow observations directly to geometry parameters.

\FloatBarrier

\subsection{Forward Operator}
\label{sec:forward_operator}
The forward problem consists of computing the steady compressible flow through a variable-area nozzle for a given geometry. Let $A(x)$ denote the cross-sectional area defined on the axial domain $x\in[0,L]$.
The forward model maps this geometry $A(x)$ to the corresponding Mach number distribution $M(x)$, governed by the quasi--one-dimensional compressible Euler equations. 

All variables in this study are expressed in nondimensional form to improve numerical conditioning and to ensure consistency across forward simulation, surrogate training, and inverse inference. The axial coordinate $x$ represents the nondimensional length of the nozzle ($L = 10$), obtained by normalizing the physical axial coordinate by the total length of the nozzle. The cross-sectional area of the nozzle $A$ is normalized to the inlet area. Density $\rho$, velocity $u$ and pressure $P$ are normalized using their corresponding inlet reference values, with velocity scaled by the square root of the inlet pressure divided by the inlet density. For clarity, nondimensional quantities are written without additional notation throughout the manuscript. \BT{Although the quasi--one-dimensional Euler equations are formulated in terms of density, velocity, and pressure, these variables are computed internally by the CFD solver as part of the forward solution process. They act as intermediate state variables and are not the primary quantities of interest in the inverse problem. In this study, the inverse formulation targets the Mach number distribution as the observable quantity used for geometry reconstruction, while density, velocity, and pressure are evaluated only to obtain the Mach field.}

This forward mapping defines a nonlinear forward operator
\begin{equation}
F:\; A(x) \longmapsto M(x).
\end{equation}
The operator $F$ is evaluated using a finite-volume CFD solver that resolves the steady nozzle flow, including supersonic acceleration in the diverging section, a standing normal shock that produces an abrupt transition from supersonic to subsonic flow, and downstream subsonic deceleration when present. The presence of shock-induced regime changes causes the Mach field to become highly sensitive to localized geometric perturbations. Even minor variations in nozzle shape can alter the position and intensity of the shock, directly affecting inverse conditioning and posterior uncertainty. The governing equations, non-dimensionalization, numerical discretization, and boundary conditions associated with this solver are detailed in Appendix~\ref{secA1}. This CFD-based evaluation of $F$ is used both to generate training data and to provide a reference solution for surrogate-accelerated and direct data-driven inverse formulation.

\subsection{Geometry Parameterization}
\label{sec:geometry_parameterization}

To obtain a finite-dimensional representation of the inverse problem, the nozzle geometry is represented as

\begin{equation}
A(x) = A(x;\boldsymbol{\theta}),
\end{equation}
where $\boldsymbol{\theta} \in \mathbb{R}^d$ denotes the vector of geometry parameters.

Three different parameterization strategies are examined. These are (i) discrete piecewise-linear, (ii) a polynomial, and (iii) cubic B-spline representations. All parameterizations enforce exact endpoint value and slope constraints to promote physically admissible nozzle geometries, and these details are provided comprehensively in the Appendix~\ref{app:geom_param}.

A preliminary CFD-based Bayesian analysis is performed to assess the impact of the geometry representation on posterior robustness with respect to parameterization and reconstruction accuracy. The analysis and comparison of the three different geometric parameterizations are presented in the results section. This comparison shows that the cubic B-spline formulation achieves the correct posterior behavior while maintaining an effective trade-off between smoothness and local control compared to the polynomial and discrete representation. Based on these findings, cubic B-spline parameterization is adopted for all surrogate model training and subsequent surrogate-accelerated inverse inference. Detailed formulations of the three geometry parameterizations are provided in the Appendix~\ref{app:geom_param}. The corresponding parameter-space priors and their induced admissible geometry envelopes are described in the Appendix~\ref{app:priors}.

\subsection{Inverse Problem Statement}
\label{sec:inverse_statement}

The inverse problem aims to recover the unknown shape of the nozzle $A(x)$ using Mach number
observations that are contaminated by noise. Let measurements or observations be available at sensor locations $\{x_j\}_{j=1}^{N_{\mathrm{obs}}}$ and modeled as
\begin{equation}
M_j^{\mathrm{obs}} = M(x_j) + \varepsilon_j,
\qquad
\varepsilon_j \sim \mathcal{N}(0,\sigma^2),
\label{eq:obs_model_results}
\end{equation}
where $\varepsilon_j$ are independent Gaussian noise terms and $\sigma^2$ denotes the variance of the observational noise. \BT{In our numerical experiments, the synthetic observations are generated using a true noise standard deviation of $\sigma_{true} = 0.03$ for $M^{obs}$. However, during Bayesian inference, $\sigma$ is treated as an unknown parameter and estimated jointly with the geometry parameters.}

The inverse problem is ill-posed due to the strong nonlinearity of the forward mapping and the limited observational information, leading to non-uniqueness in the recovered geometry. A Bayesian formulation is therefore adopted to characterize posterior uncertainty instead of producing a single deterministic estimate.

\subsection{Bayesian Formulation}
\label{sec:bayesian_formulation}

Bayesian inference is used to infer the posterior distribution of the geometry parameters and noise scale.
For a given parameter vector $\boldsymbol{\theta}$, the predicted Mach values at the observation locations are
\begin{equation}
M_j(\boldsymbol{\theta}) = \big(F(A(\cdot;\boldsymbol{\theta}))\big)(x_j).
\end{equation}
Under the Gaussian noise model, the likelihood function is
\begin{equation}
p(\mathbf{M}^{\mathrm{obs}} \mid \boldsymbol{\theta},\sigma^2)
\propto
\exp\left(
-\frac{1}{2\sigma^2}
\sum_{j=1}^{N_{\mathrm{obs}}}
\big(M_j^{\mathrm{obs}} - M_j(\boldsymbol{\theta})\big)^2
\right).
\end{equation}
Prior distributions are imposed to encode physical regularity and scale information,
\begin{equation}
\boldsymbol{\theta} \sim p(\boldsymbol{\theta}),
\qquad
\sigma \sim \mathrm{HalfNormal}(\sigma_0),
\end{equation}
where $p(\boldsymbol{\theta})$ enforces smooth, physically admissible geometries and \BT{$\sigma$ denotes the observational noise standard deviation, which is treated as an unknown parameter inferred jointly with the geometry parameters. A HalfNormal prior is used for $\sigma$ to enforce the positivity of the noise scale while providing a weakly informative prior commonly used for variance parameters in Bayesian models. The hyperparameter $\sigma_0$ controls the scale of this prior and reflects the expected magnitude of the observational noise. In the numerical experiments considered in this work, the observations are generated by adding Gaussian noise with standard deviation $\sigma_{\mathrm{true}} = 0.03$ to the true Mach number distribution. Accordingly, we choose $\sigma_0 = 0.05$, which is slightly larger than the prescribed observational noise level so that the prior remains weakly informative while allowing the noise scale to be inferred flexibly from the data.} The specific prior constructions for each geometry parameterization are detailed in Appendix~\ref{app:priors}. The posterior distribution is
then given by Bayes' theorem~\cite{gelman1995bayesian},
\begin{equation}
p(\boldsymbol{\theta},\sigma \mid \mathbf{M}^{\mathrm{obs}})
\propto
p(\mathbf{M}^{\mathrm{obs}} \mid \boldsymbol{\theta},\sigma^2)\,
p(\boldsymbol{\theta})\,
p(\sigma).
\end{equation}

\subsubsection{Hamiltonian Monte Carlo and NUTS}
\label{sec:hmc_nuts}

\BT{To simultaneously infer the geometry parameters and the observational noise scale, let $\boldsymbol{\Theta} = \{\boldsymbol{\theta}, \sigma\}$ denote the joint parameter vector.} The posterior distribution is explored using Hamiltonian Monte Carlo (HMC), which introduces an auxiliary momentum variable $r$ of the same dimension as $\boldsymbol{\Theta}$ and defines the Hamiltonian
\begin{equation}
H(\boldsymbol{\Theta},r)
=
-\log p(\boldsymbol{\Theta} \mid \mathbf{M}^{\mathrm{obs}})
+
\frac{1}{2} r^\top \mathcal{M}^{-1} r,
\end{equation}
where $\mathcal{M}$ is a mass matrix. HMC proposals are generated by simulating Hamiltonian dynamics across the joint parameter space,
\begin{equation}
\frac{d\boldsymbol{\Theta}}{dt} = \mathcal{M}^{-1} r,
\qquad
\frac{dr}{dt} = \nabla_{\boldsymbol{\Theta}} \log p(\boldsymbol{\Theta} \mid \mathbf{M}^{\mathrm{obs}}),
\end{equation}
using a symplectic integrator.

To eliminate manual tuning of trajectory length, the No-U-Turn Sampler (NUTS) is employed.
NUTS adaptively constructs Hamiltonian trajectories and terminates integration when a reversal in phase-space motion is detected. All gradients required by NUTS are computed via automatic differentiation. In the surrogate-accelerated pipeline, the CFD operator within the likelihood is
replaced by the differentiable DeepONet surrogate, yielding orders-of-magnitude reductions in
computational cost while preserving posterior structure.




\subsection{Neural Operator Surrogates}
\label{sec:neural_operator_surrogates}

\subsubsection{Forward Neural Operator}
\label{sec:forward_neural_operator}

To alleviate the computational cost associated with repeated evaluations of the CFD solver during Bayesian sampling, a Deep Operator Network (DeepONet) is trained offline to approximate the forward mapping
\begin{equation}
\widehat{F} : A(x) \longmapsto \widehat{M}(x).
\end{equation}

The forward DeepONet consists of a branch network that encodes the nozzle geometry and a trunk network that encodes the spatial coordinate. The Mach number prediction is obtained through a latent inner-product representation between the branch and trunk outputs. The surrogate is trained using CFD-generated geometry--flow pairs sampled from the cubic B-spline geometry parameterization. Details on network architecture, training procedure, and validation diagnostics are provided in the Appendix~\ref{sec:deeponet_appendix}.

Once trained, the forward DeepONet is frozen and substituted directly into the Bayesian likelihood function. Importantly, the Bayesian formulation itself, including the likelihood, priors, and sampling strategy, remains unchanged. This substitution isolates the impact of the surrogate approximation error while enabling orders-of-magnitude reductions in computational cost.

\subsubsection{Direct Inverse Neural Operator}
\label{sec:inverse_neural_operator}

In addition to Bayesian inversion, we consider a deterministic alternative based on learning the inverse mapping directly. An inverse DeepONet is trained to approximate the operator
\begin{equation}
\widehat{G}:\; \big(M_{\mathrm{sparse}}(x),\, \omega(x)\big) \longmapsto \widehat{A}(x),
\end{equation}
where $M_{\mathrm{sparse}}(x)$ denotes a sparsely observed Mach field and $\omega(x)\in\{0,1\}$ is a binary observation mask indicating the spatial locations of available sensors.

Sparse observations are embedded on the full spatial grid by defining a vector $M_{\mathrm{sparse}}(x)\in\mathbb{R}^N$ such that $M_{\mathrm{sparse}}(x_i)=M_j^{\mathrm{obs}}\text{ at sensor locations } x_i$ and $M_{\mathrm{sparse}}(x_i)=0$ elsewhere. The corresponding mask satisfies $\omega(x_i)=1$ if a sensor is present at $x_i$ and $\omega(x_i)=0$ otherwise. This construction preserves a fixed input dimension while explicitly distinguishing observed from unobserved locations.

Although $M_{\mathrm{sparse}}(x)$ and $\omega(x)$ are provided as separate inputs to the network, the inverse DeepONet learns during training to suppress contributions from unobserved locations. Because Mach values associated with $\omega(x_i)=0$ are consistently non-informative across randomized sparsification patterns, gradient-based optimization drives the network to rely on $M_{\mathrm{sparse}}(x_i)$ only when $\omega(x_i)=1$, effectively learning a soft gating mechanism without explicitly multiplying the two inputs.

The inverse DeepONet adopts the same operator-learning structure as the forward surrogate, consisting of a branch network and a trunk network coupled through a latent inner-product representation. The branch network encodes the concatenated input vector
\begin{equation}
\mathbf{z} = \big[ M_{\mathrm{sparse}}(x),\, \omega(x) \big] \in \mathbb{R}^{2N},
\end{equation}
while the trunk network encodes the spatial coordinate $x$. The reconstructed geometry field is obtained pointwise as
\begin{equation}
\widehat{A}(x_i) = \sum_{k=1}^{p} b_k\!\left(M_{\mathrm{sparse}}, \omega\right)\, t_k(x_i) + b_0,
\end{equation}
where $b_k$ are geometry-dependent coefficients produced by the branch network, $t_k(x)$ are spatial basis functions produced by the trunk network, $p$ is the latent dimension, and $b_0$ is a scalar bias.

To promote robustness with respect to sensor placement and sensor density, sparsification is applied dynamically during training. For each training sample, both the number of sensors and their spatial arrangement are randomly selected. Three sensor placement strategies are considered: uniform, jittered-uniform, and random. Jittered-uniform layouts are constructed by first placing sensors at uniformly spaced locations and then perturbing interior sensor positions by a small fraction of the local spacing while keeping the domain endpoints fixed. This preserves global spatial coverage while introducing controlled irregularity, preventing the network from overfitting to perfectly regular sensor grids.

\BT{It is important to note that the inverse mapping from Mach number fields to nozzle geometries may not be strictly one-to-one. In general, multiple geometries may produce similar Mach distributions, particularly when observations are sparse. The inverse DeepONet does not explicitly model this non-uniqueness. Instead, during training the network learns a deterministic mapping determined by its optimized weights and biases using pairs of geometries and their corresponding Mach fields generated by the forward solver. Consequently, when a sparse Mach field is provided as input, the trained network produces a single geometry prediction corresponding to a representative solution within the geometry manifold learned from the training data. Although the present study focuses on deterministic reconstruction using DeepONet, extensions that explicitly account for non-uniqueness, such as probabilistic neural operators or Bayesian learning frameworks, could be explored in future work to represent multiple admissible geometries consistent with the observed Mach field.}

Although the forward and inverse neural operators share a common DeepONet architecture, they serve fundamentally different roles. The forward surrogate is embedded within a probabilistic Bayesian inference framework and must be accurate and differentiable over the posterior support. In contrast, the inverse operator performs a single-shot deterministic reconstruction and does not provide uncertainty quantification. The inverse DeepONet is therefore presented as an alternative to surrogate-accelerated Bayesian inversion rather than replacing the forward model within a probabilistic framework. It bypasses Bayesian inference entirely by learning a deterministic mapping from sparse flow observations to geometry. The inverse DeepONet is trained using the same architecture, normalization strategy, and optimization procedure described in Appendix~\ref{sec:deeponet_appendix}, with the training objective adapted to geometry reconstruction.

\subsection{Evaluation Metrics}
\label{sec:evaluation_metrics}

The accuracy of the Bayesian reconstructions is quantified using the posterior mean of the inferred
fields. For a generic field $f(x)$, representing either the nozzle geometry $A(x)$ or the Mach
distribution $M(x)$, the posterior mean at spatial location $x_i$ is defined as
\begin{equation}
\overline{f}(x_i)
=
\frac{1}{N_s}
\sum_{k=1}^{N_s} f^{(k)}(x_i),
\end{equation}
where $\{f^{(k)}\}_{k=1}^{N_s}$ denotes the retained posterior samples after burn-in.

Reconstruction accuracy is measured using the root-mean-square error (RMSE) of the posterior mean
relative to a reference solution $f_{\mathrm{ref}}(x)$,
\begin{equation}
\mathrm{RMSE}(f)
=
\left(
\frac{1}{N}
\sum_{i=1}^{N}
\bigl(
\overline{f}(x_i) - f_{\mathrm{ref}}(x_i)
\bigr)^2
\right)^{1/2},
\end{equation}
where $N$ is the number of spatial grid points. \BT{The posterior mean is used as the primary estimator for quantitative accuracy because it provides a single representative reconstruction that can be directly compared with the reference solution using deterministic error metrics such as RMSE. While Bayesian inference produces a full posterior distribution, the posterior mean enables consistent comparison of reconstruction accuracy across different forward models and inference configurations. Posterior dispersion is assessed through 
95 $\%$ credible intervals computed from empirical quantiles of the sampled fields. These intervals are presented in the results to illustrate the spatial structure and contraction behavior of the posterior uncertainty across inference settings.}

To distinguish between forward models, we denote the RMSE of the geometry reconstruction obtained
using the CFD-based forward solver as $\mathrm{RMSE}_A^{\mathrm{CFD}}$, and the corresponding error
obtained using the DeepONet surrogate as $\mathrm{RMSE}_A^{\mathrm{DeepONet}}$. Analogous notation is
used for the Mach field, $\mathrm{RMSE}_M^{\mathrm{CFD}}$ and
$\mathrm{RMSE}_M^{\mathrm{DeepONet}}$.

For quantitative comparison, we additionally report the ratio of surrogate-based error to
CFD-based error,
\begin{equation}
\mathrm{Ratio}_f
=
\frac{\mathrm{RMSE}_f^{\mathrm{DeepONet}}}
     {\mathrm{RMSE}_f^{\mathrm{CFD}}},
\qquad
f \in \{A, M\},
\label{eq:rmse_ratio}
\end{equation}
which provides a direct measure of how closely the surrogate-accelerated inference reproduces the
accuracy of the CFD-based reference.

Computational efficiency is quantified using the total wall-clock time required to complete Bayesian
inference with NUTS. Let $T_{\mathrm{CFD}}$ and $T_{\mathrm{DeepONet}}$ denote the total inference time
when using the CFD solver and the DeepONet surrogate, respectively. The resulting acceleration is
measured through the speedup factor
\begin{equation}
\mathrm{Speedup}
=
\frac{T_{\mathrm{CFD}}}{T_{\mathrm{DeepONet}}},
\end{equation}
which captures the practical computational benefit of replacing the high-fidelity solver with a
neural operator within the Bayesian inference loop.

\section{Results}
\label{sec:results}

This section presents numerical results for the Bayesian inverse problem of reconstructing the
nozzle area distribution $A(x)$ from Mach number observations with varying sensor density.
We begin by defining the reference problem and observation scenarios used throughout the study.
A CFD-based Bayesian inverse formulation is then used to establish a high-fidelity reference and to
examine the influence of geometry parameterization on posterior stability and reconstruction accuracy.
Next, a DeepONet forward surrogate is validated and embedded within the Bayesian inference framework
to enable surrogate-accelerated posterior sampling, with reconstruction accuracy and uncertainty
compared directly against the CFD-based reference across sparse to fully observed regimes.
Computational efficiency and speedup are subsequently quantified.
Finally, a direct inverse neural operator is examined as a deterministic alternative for inverse
design, providing qualitative comparison with Bayesian posterior means while highlighting the role
of uncertainty-aware inference.

\subsection{Reference Problem Setup and Observation Scenarios}
\label{sec:results_setup}

\begin{figure}[H]
  \centering
  \includegraphics[width=0.98\textwidth]{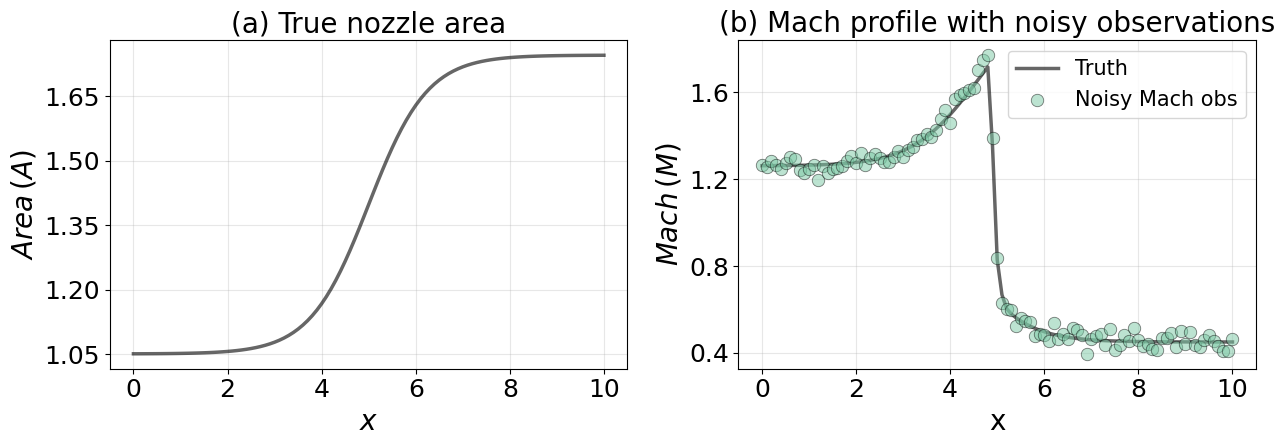}
  \caption{Reference problem definition.
  (a) Ground-truth nozzle area distribution $A_{\mathrm{true}}(x)$.
  (b) Corresponding Mach number field $M_{\mathrm{true}}(x)$ obtained from the quasi--1D Euler solver
  (solid line) together with synthetic noisy observations generated using
  Eq.~\eqref{eq:obs_model_results} with $\sigma=0.03$ (green markers).}
  \label{fig:R1_setup}
\end{figure}

Fig.~\ref{fig:R1_setup} defines the reference forward problem used throughout this study.
Fig.~\ref{fig:R1_setup}(a) shows the ground-truth nozzle area distribution $A_{\mathrm{true}}(x)$, which serves as
the target geometry for all inverse reconstructions. Fig.~\ref{fig:R1_setup}(b) shows the corresponding Mach number
field $M_{\mathrm{true}}(x)=F(A_{\mathrm{true}}(x))$ obtained from the quasi--one-dimensional Euler
solver as described in methods and Appendix~\ref{secA1}. The reference Mach field $M_{\mathrm{true}}(x)$ is evaluated on the full
$N = 101$-point CFD grid.
The flow accelerates in the supersonic regime within the diverging nozzle and undergoes a standing normal shock, producing a discontinuous transition to subsonic flow downstream. This shock-induced nonlinearity leads to a highly sensitive and non-smooth mapping between the nozzle geometry and the resulting flow field.

Synthetic observations are generated by sampling the reference Mach field at discrete axial locations
and contaminating the samples with additive Gaussian noise from the interior point of the true field. In all numerical experiments, a fixed noise level corresponding to the relative uncertainty of $3\%$ in the Mach number is used, representing moderately noisy experimental or flight-derived measurements.

\FloatBarrier

\begin{figure}[H]
  \centering
  \includegraphics[width=0.98\textwidth]{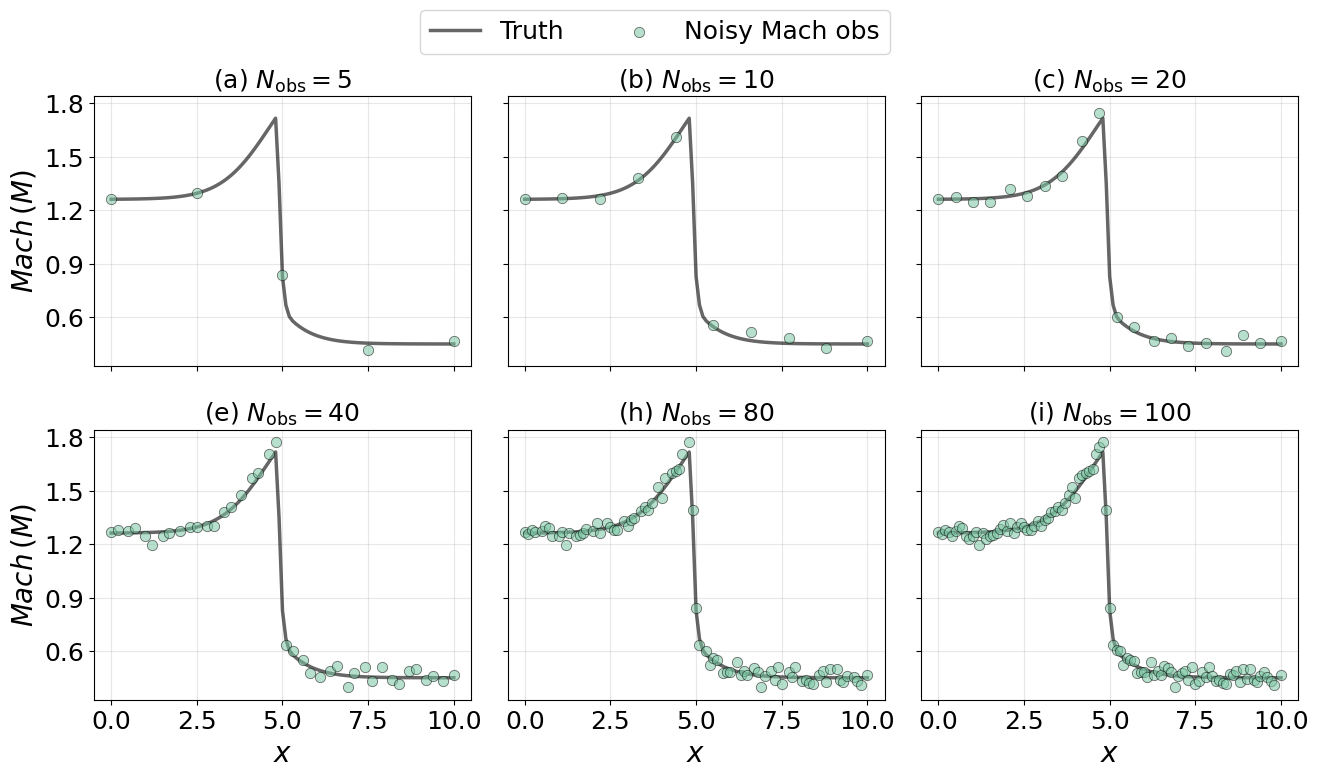}
 \caption{Mach number observation scenarios for varying sensor densities.
Noisy Mach observations (green markers) are shown together with the reference solution (solid line) for
(a) $N_{\mathrm{obs}}=5$,
(b) $N_{\mathrm{obs}}=10$,
(c) $N_{\mathrm{obs}}=20$,
(d) $N_{\mathrm{obs}}=40$,
(e) $N_{\mathrm{obs}}=80$,
and (f) $N_{\mathrm{obs}}=100$.
Increasing observation density progressively resolves flow characteristic and the shock-induced discontinuity.}

  \label{fig:R2_obs_cases}
\end{figure}

To systematically assess the impact of data availability on the inverse problem, multiple observation
densities are considered, as summarized in Fig.~\ref{fig:R2_obs_cases}. The number of observation points ranges from very sparse to fully resolved, with sensor counts of $ N_{obs} = 5, 10, 20, 40, 80, \text{and}\,100$ uniformly distributed throughout the domain. For highly sparse cases (5 and 10 observations), the measurements capture only the global trend of the Mach profile, while the shock location and strength are poorly constrained. The case with 20 observations represents a sparse but practically relevant regime, in which only a small fraction of the domain is observed and the shock is weakly resolved. As the number of observations increases to 40 and 80, the shock structure and post-shock behavior become progressively better resolved, although sensitivity near the discontinuity remains limited. The fully observed case with $N_{\mathrm{obs}}=100$  corresponds to observations at
all interior grid points and results in a comparatively well-conditioned inverse problem.

\FloatBarrier

Figures~\ref{fig:R1_setup} and~\ref{fig:R2_obs_cases} together highlight the intrinsic difficulty of the
nozzle inverse problem. The forward mapping $A(x)\mapsto M(x)$ is highly nonlinear in the rapid area-transition region
($x \approx 4.5$--$5.5$) and in the downstream shock-affected region
($x \approx 5.5$--$6.5$), such that small geometric perturbations can induce large localized changes in the observations. As the number of observations decreases, the problem becomes increasingly ill posed, with multiple
geometries consistent with the same sparse and noisy data. These characteristics motivate the Bayesian
formulation adopted in this work, which incorporates prior regularization and yields posterior
uncertainty quantification rather than a single deterministic reconstruction.

\subsection{Effect of Geometry Parameterization}
\label{sec:results_cfd_inverse}

\begin{figure}[H]
  \centering
  \includegraphics[width=0.9\textwidth]{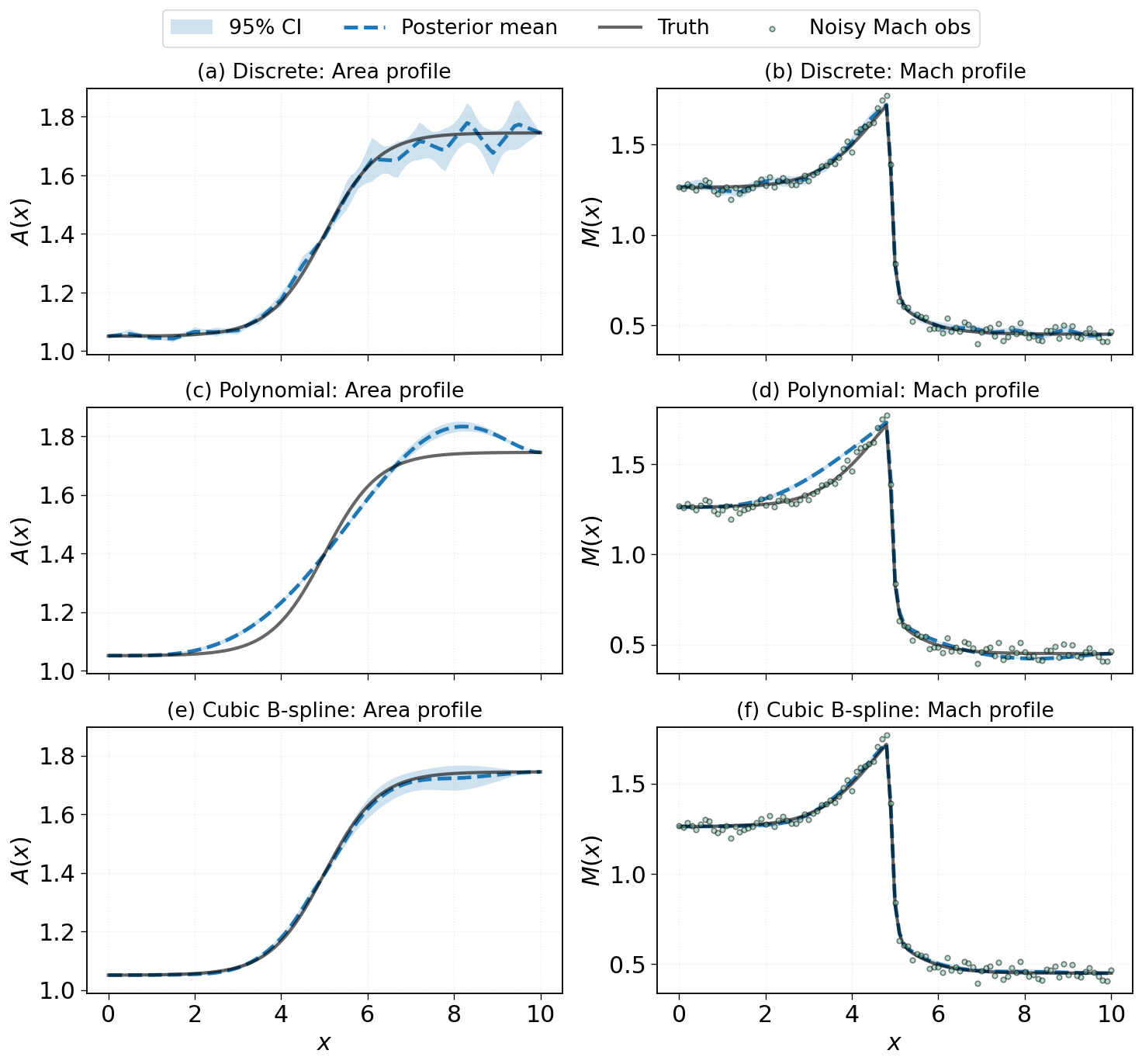}
  \caption{CFD-based Bayesian posterior reconstructions under full observation ($N_{\mathrm{obs}}=100$) for
  three geometry parameterizations. Left column: inferred nozzle area $A(x)$ (posterior mean and $95\%$ credible
  interval) versus the reference geometry. Right column: posterior predictive Mach field $M(x)$ versus the CFD
  truth, with noisy Mach observations overlaid.}
  \label{fig:R3_param_compare}
\end{figure}

We first solve the Bayesian inverse problem using the full CFD solver embedded within NUTS to obtain a
high-fidelity reference posterior. To isolate the influence of geometry representation from data
sparsity, all results in this subsection use the fully observed setting ($N_{\mathrm{obs}}=100$) with $\sigma_{true}=0.03$ noise level. Fig.~\ref{fig:R3_param_compare} summarizes posterior
reconstructions of the nozzle area $A(x)$ and the posterior predictive Mach profile $M(x)$, where the
noisy Mach observations are also shown.

Although the reconstructed Mach field closely matches the CFD solution across the domain, the inferred geometry exhibits elevated uncertainty downstream of the shock ($x \approx 6$--$10$), where the Mach field is only weakly sensitive to local geometric variations. This reflects the reduced sensitivity of subsonic flow to geometric perturbations in the diffuser, where multiple geometries produce nearly indistinguishable Mach distributions, leading to intrinsic non-identifiability. We then compare three geometry parameterizations under
identical endpoint value and slope constraints: a discrete piecewise-linear representation, a global
seventh-degree polynomial, and a cubic B-spline representation. The discrete model exhibits visibly
larger credible intervals and mild localized oscillations in $A(x)$, reflecting limited smoothness.
The polynomial model is smooth but shows broader uncertainty due to global coefficient coupling, which
reduces local identifiability. In contrast, the cubic B-spline representation yields the most stable
posterior, characterized by narrow credible intervals and minimal spurious oscillations, consistent with its smoothness and strong local control. These trends are consistent with the structure of the underlying priors, whose induced admissible geometry envelopes are shown in Fig.~\ref{fig:prior_comparison} of Appendix~\ref{app:priors}.

\begin{figure}[H]
  \centering
  \includegraphics[width=0.60\textwidth]{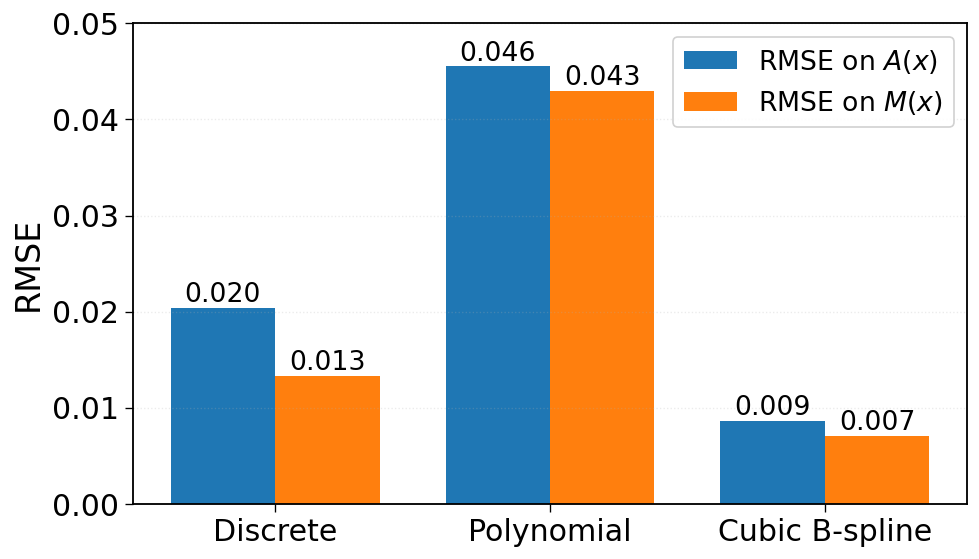}
  \caption{RMSE of the posterior mean for geometry ($A(x)$) and posterior predictive Mach field ($M(x)$) under
  full observation ($N_{\mathrm{obs}}=100$) for each geometry parameterization.}
  \label{fig:R3_rmse}
\end{figure}

These qualitative trends are confirmed by the quantitative errors in Fig.~\ref{fig:R3_rmse}. The
cubic B-spline achieves the lowest reconstruction errors for both the geometry and flow: RMSE
$=0.009$ for $A(x)$ and RMSE $=0.007$ for $M(x)$, compared to the discrete case (RMSE $=0.020$ for
$A(x)$ and $0.013$ for $M(x)$) and the polynomial case (RMSE $=0.046$ for $A(x)$ and $0.043$ for
$M(x)$). Based on this CFD-based study, the cubic B-spline parameterization is selected as the
reference geometry model for subsequent surrogate training and surrogate-accelerated Bayesian
inference, as it provides the best balance of smoothness, local control, and posterior conditioning.

\FloatBarrier

\FloatBarrier

\subsection{Neural Operator Surrogates: Forward Model Validation}
\label{sec:forward_surrogate_validation}

Before employing the neural operator within Bayesian inference, we first assess the accuracy and generalization capability of the forward DeepONet independently of any inverse task. The surrogate is trained to approximate the nonlinear operator mapping nozzle geometry to the Mach number field using CFD-generated data sampled from the cubic B-spline parameterization.

\begin{figure}[H]
  \centering
  \includegraphics[width=0.95\textwidth]{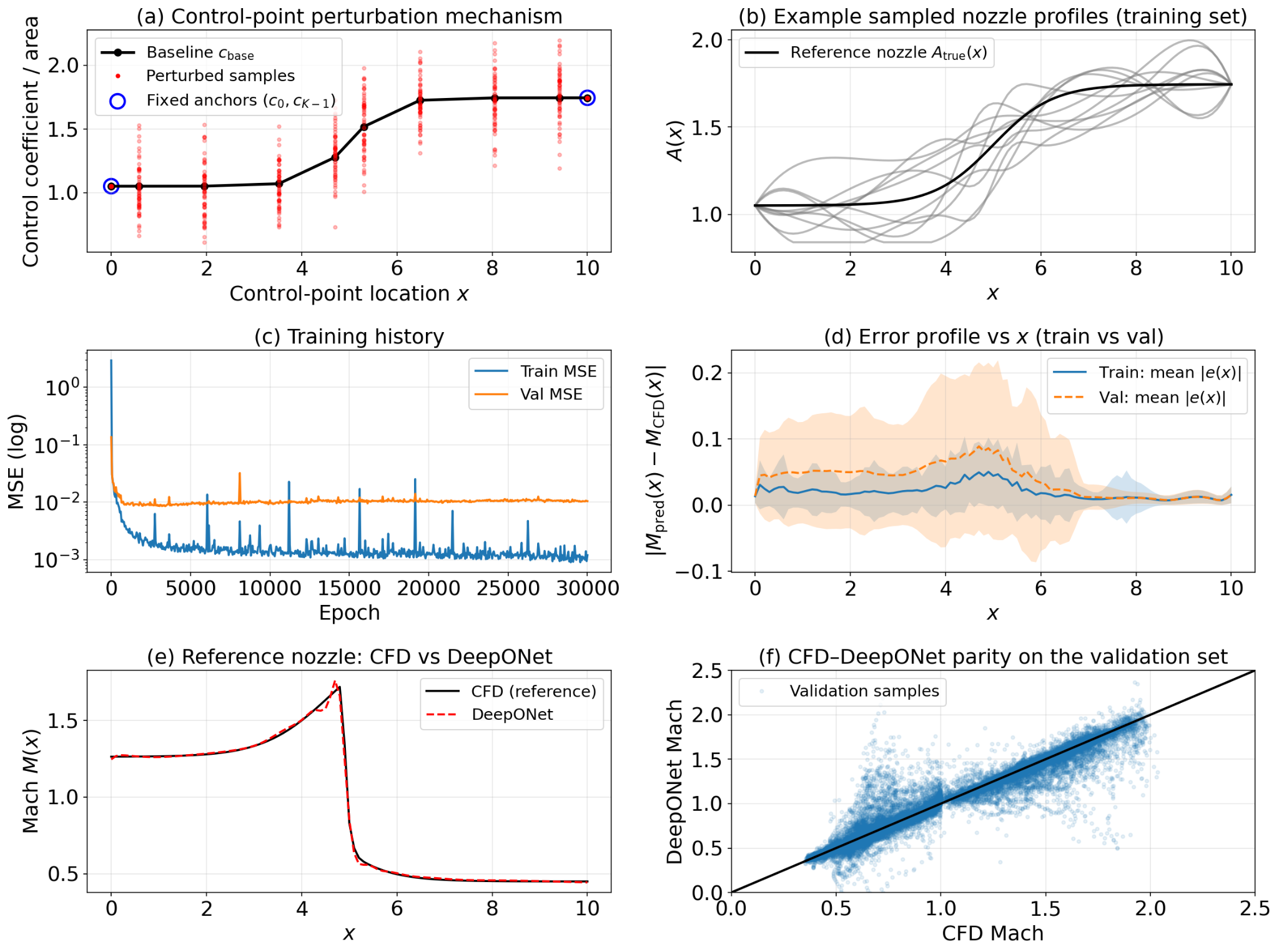}
  \caption{DeepONet surrogate workflow and diagnostic evaluation. (a) Control-point perturbation mechanism
  used to sample cubic B-spline nozzle geometries while enforcing fixed endpoint constraints. (b) Example
  sampled nozzle area profiles from the training set overlaid with the reference nozzle \(A_{\mathrm{true}}(x)\). \BT{While additional structural constraints, such as monotonicity or other physically motivated regularity conditions, could enforce more realistic area distributions, we deliberately retain an unconstrained parameterization of the control-point perturbations to avoid embedding prior geometric assumptions and to permit an unbiased exploration of the admissible functional space.}
  (c) Training history showing the mean-squared error (MSE) on the training and validation sets.
  (d) Mean absolute error profile \({e}(x)\) as a function of
  axial position, shown for train and validation sets; shaded regions denote standard deviation.
  (e) Reference-nozzle comparison between CFD and DeepONet. (f) Validation parity plot over all validation
  samples and spatial grid points; the point lying on the diagonal indicates perfect agreement.}
  \label{fig:appendixC_deeponet}
\end{figure}

Fig.~\ref{fig:appendixC_deeponet} summarizes the validation results. The training and validation loss histories indicate stable convergence without evidence of overfitting. Spatial error profiles show uniformly small prediction errors across the domain, with modest error amplification localized in the rapid area-transition region
($x \approx 4.5$--$5.5$) and in the downstream shock-affected region
($x \approx 5.5$--$6.5$), where the flow response is most sensitive to geometric perturbations. Representative comparisons between CFD and DeepONet predictions demonstrate accurate reproduction of both smooth acceleration and the shock-induced Mach discontinuity. The parity plot further confirms strong pointwise agreement between surrogate predictions and CFD solutions across the full validation dataset.

These results indicate that the forward DeepONet provides an accurate and stable approximation of the CFD operator over the geometry distribution used for training and validation. Nevertheless, the surrogate remains an approximation and may introduce localized errors, particularly in regions of strong nonlinearity such as the rapid transition associated with shock formation and the downstream region. When embedded within the inverse or Bayesian inference framework, such approximation errors can propagate into the reconstructed geometry and posterior uncertainty. For this reason, the present validation establishes surrogate reliability but does not eliminate surrogate-induced bias, which is explicitly assessed in subsequent sections through direct comparison between CFD-based and surrogate-accelerated Bayesian inference.

\FloatBarrier

\subsection{Surrogate-Accelerated Bayesian Inference}
\label{sec:results_surrogate_inverse}

Having established a CFD-based reference posterior and independently validated the DeepONet forward
operator as described in previous section, we next replace the CFD solver within the Bayesian
likelihood evaluation by the trained DeepONet surrogate. All remaining components of the inference
pipeline, including the cubic B-spline geometry parameterization, prior distributions, observation
model, and NUTS sampling configuration, are held fixed. This controlled substitution isolates the
impact of surrogate-based acceleration on posterior accuracy and uncertainty.

\begin{figure}[H]
  \centering
  \includegraphics[width=\linewidth]{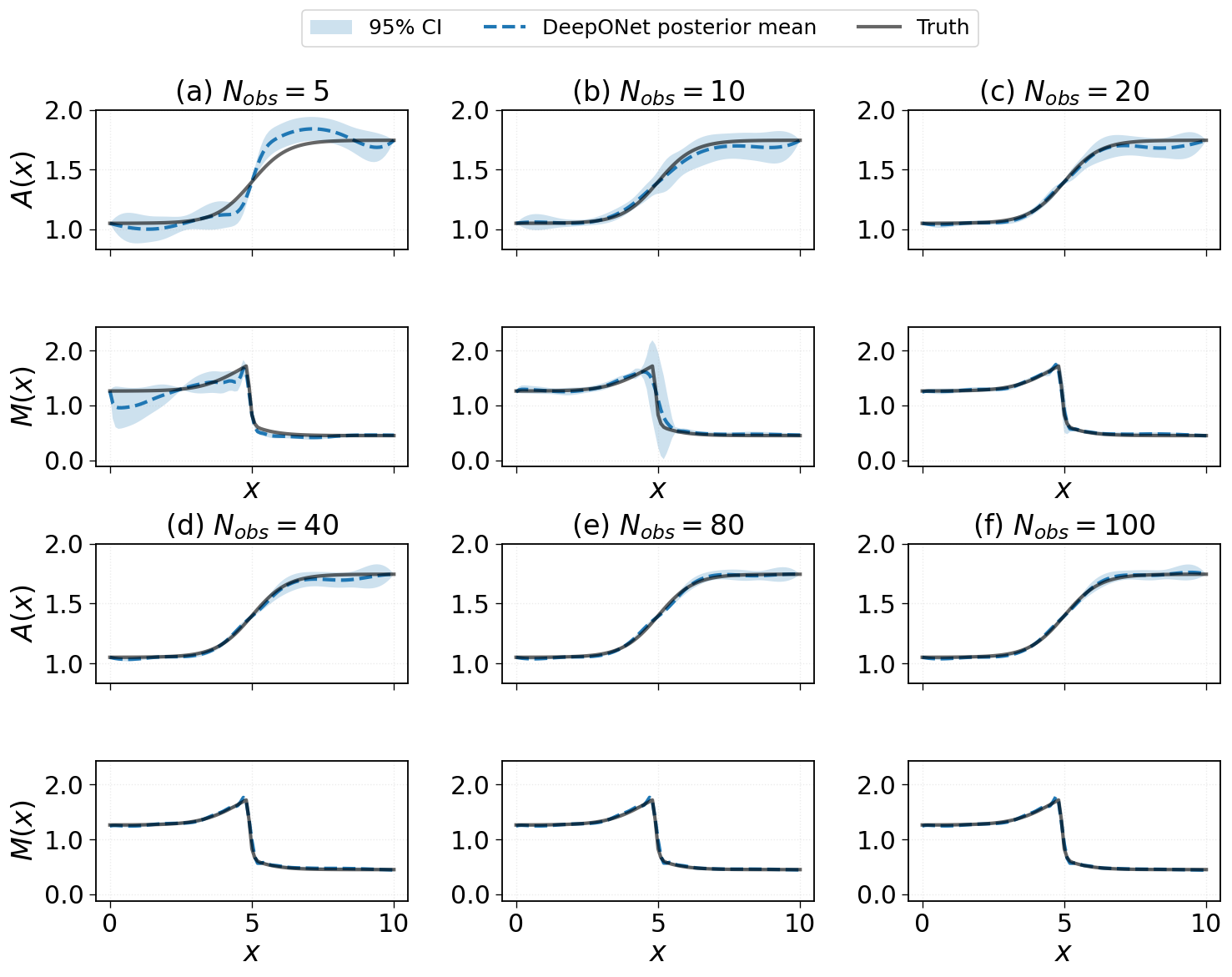}
  \caption{DeepONet-based Bayesian posterior reconstructions of the nozzle geometry $A(x)$
(rows 1 and 3) and Mach field $M(x)$ (rows 2 and 4) for increasing observation densities
($N_{\mathrm{obs}} = 5,10,20,40,80,100$). Solid curves denote posterior means, shaded regions indicate
$95\%$ credible intervals, and black curves denote the true solution.}
  \label{fig:R5}
\end{figure}

Fig.~\ref{fig:R5} presents DeepONet-based posterior reconstructions of the nozzle geometry $A(x)$
and Mach field $M(x)$ for increasing observation densities
($N_{\mathrm{obs}} = 5,10,20,40,80,$ and $100$).
For the geometry reconstruction, the posterior mean closely follows the true nozzle shape across
all cases, accurately recovering the smooth area variation and the rapid transition region associated with shock formation. As the number of observations increases, the $95\%$ credible intervals contract monotonically,
with the highest uncertainty localized in the rapid area-transition region
($x \approx 4.5$--$5.5$) and in the downstream shock-affected region for sparse data.
This behavior reflects the ill-posed nature of the inverse problem in the data-limited regime.

The corresponding Mach field reconstructions exhibit similarly robust behavior. The surrogate-based posterior accurately captures the upstream acceleration, the rapid transition associated with shock formation
near $x \approx 5$, and the downstream shock-induced deceleration. For sparse observation regimes
($N_{\mathrm{obs}} \le 10$), increased posterior variance persists in the vicinity of the shock,
reflecting strong local nonlinearity and sensitivity to limited measurements. For
$N_{\mathrm{obs}} \ge 40$, both the posterior mean and uncertainty envelope closely align with the
true solution throughout the domain.

\begin{figure}[H]
  \centering
  \includegraphics[width=\linewidth]{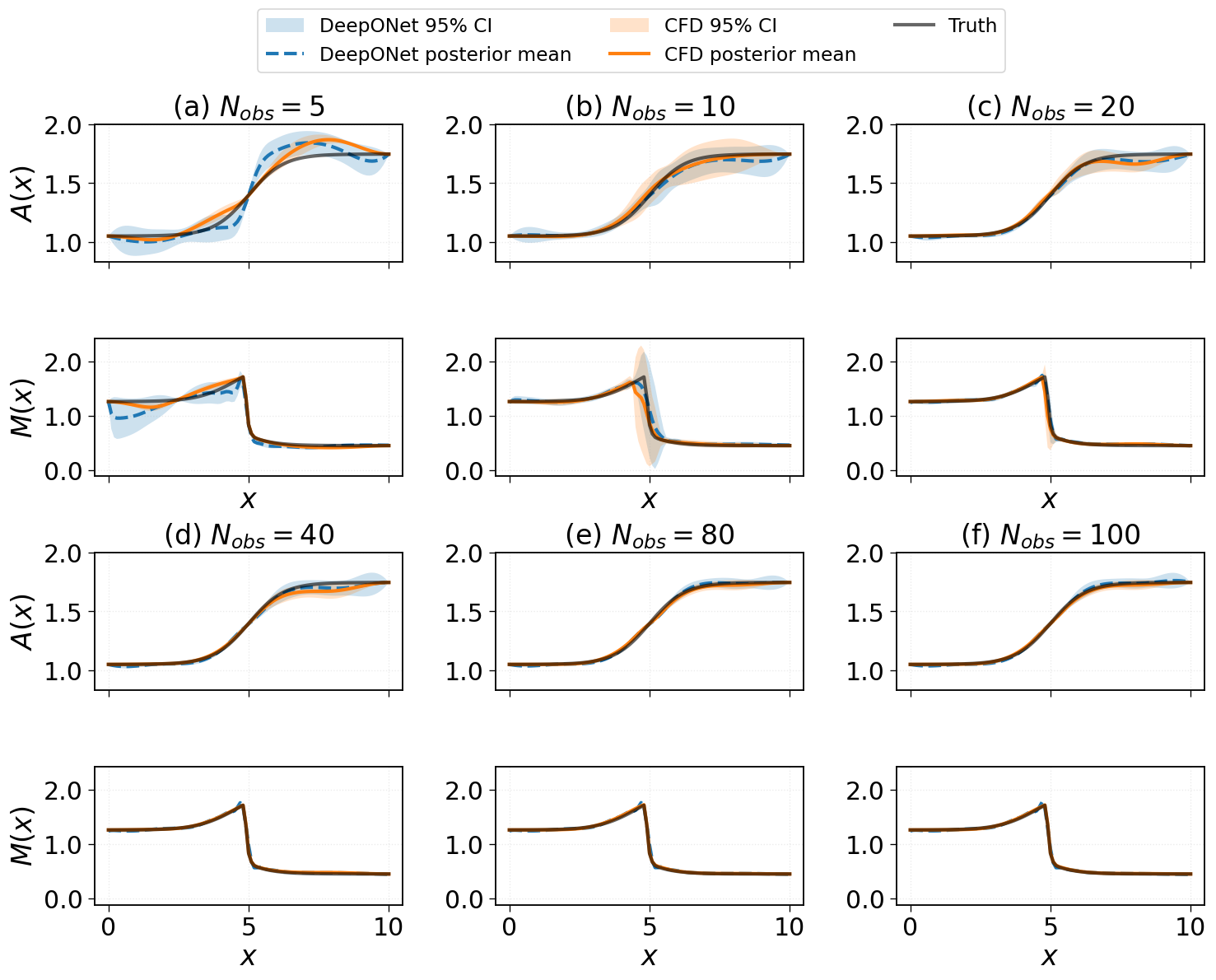}
 \caption{Comparison of CFD-based and DeepONet-based Bayesian posterior reconstructions for the nozzle
geometry $A(x)$ (rows 1 and 3) and Mach field $M(x)$ (rows 2 and 4) across observation densities
($N_{\mathrm{obs}} = 5,10,20,40,80,100$). Posterior means and $95\%$ credible intervals are shown for
both forward models, along with the true solution.}
  \label{fig:R6}
\end{figure}

To directly assess the fidelity of the surrogate-accelerated inference relative to the CFD-based
baseline, Fig.~\ref{fig:R6} compares posterior reconstructions obtained using CFD and DeepONet
forward models under identical inference settings. Across all observation densities, the posterior
means inferred using DeepONet closely match those obtained using CFD for both $A(x)$ and $M(x)$.
Moreover, the spatial structure and contraction behavior of the credible intervals are nearly
identical, with minor discrepancies confined to regions of elevated posterior variance near the
shock for sparse data.

\begin{figure}[H]
  \centering
  \includegraphics[width=0.85\linewidth]{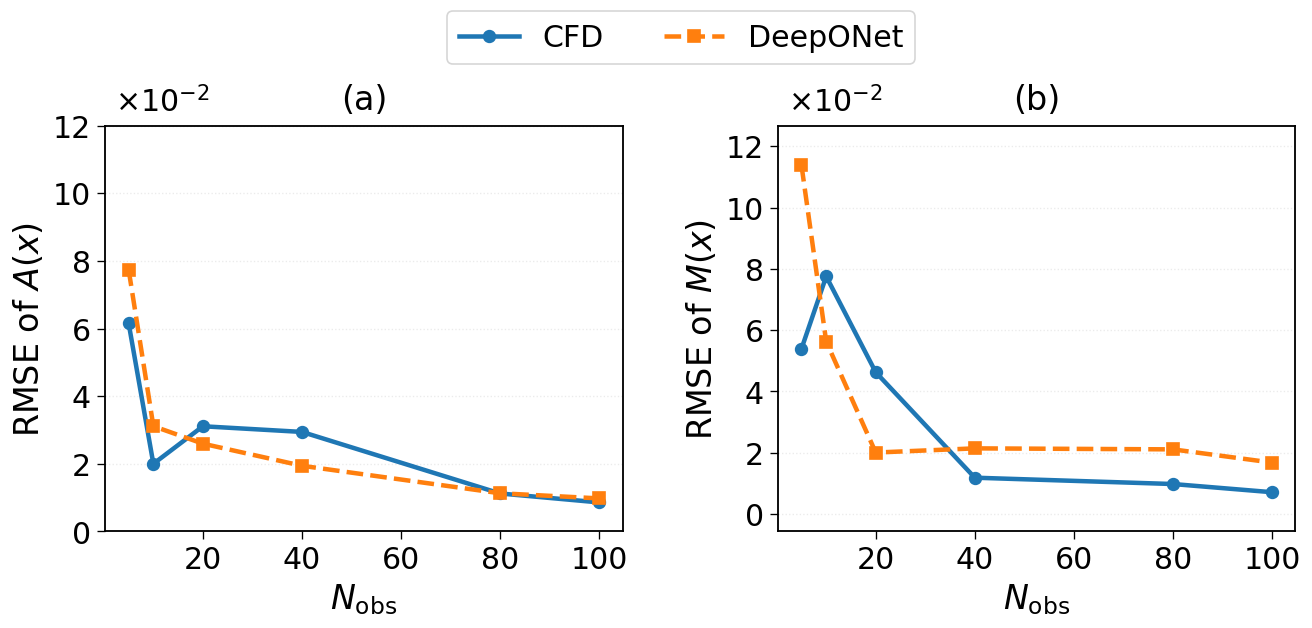}
  \caption{Quantitative comparison of posterior-mean reconstruction error as a function of
observation density $N_{\mathrm{obs}}$.
Panels report RMSE for (a) the reconstructed geometry $A(x)$ and (b) the Mach field $M(x)$ using
CFD-based and DeepONet-based Bayesian inference.
RMSE values are reported in units of $\times 10^{-2}$.}
  \label{fig:R7}
\end{figure}

A quantitative comparison of posterior-mean reconstruction accuracy is provided in
Fig.~\ref{fig:R7}. The root-mean-square error (RMSE), defined in
Section~\ref{sec:evaluation_metrics}, is computed for the posterior mean of both the reconstructed
geometry $A(x)$ and the Mach field $M(x)$ as functions of observation density.
Figures~\ref{fig:R7}(a,b) report the RMSE values obtained using CFD-based and DeepONet-based inference,
while Fig.~\ref{fig:speedup_summary}(a) shows the ratio of DeepONet RMSE to CFD RMSE.

For the geometry reconstruction, the DeepONet-based RMSE is comparable to or lower than the
CFD-based RMSE for sparse and moderate observation densities
($N_{\mathrm{obs}} \le 40$). As the number of observations increases, the geometry RMSE obtained
using DeepONet approaches that of the CFD-based inference, yielding comparable errors for
$N_{\mathrm{obs}} = 80$ and a marginally larger error at $N_{\mathrm{obs}} = 100$.

For the Mach field, the DeepONet-based RMSE is lower than the CFD-based RMSE for the sparsest cases
($N_{\mathrm{obs}} = 5$ and $10$), but exceeds the CFD-based RMSE for
$N_{\mathrm{obs}} \ge 40$. This trend reflects the increasing influence of localized shock
resolution errors in the surrogate prediction as the posterior becomes more tightly constrained by
dense observations.

In general, Figures~\ref{fig:R5}--\ref{fig:R7} and Table~\ref{tab:rmse_summary} demonstrate that replacing
the CFD solver with a DeepONet surrogate preserves the essential structure of the Bayesian posterior,
including both the posterior mean behavior and the evolution of uncertainty.

\FloatBarrier

\FloatBarrier

\FloatBarrier

\begin{table}[t]
\centering
\caption{RMSE comparison between CFD-based and DeepONet-based Bayesian inference
for reconstructed geometry $A(x)$ and predicted Mach field $M(x)$ as a function of observation
density $N_{\mathrm{obs}}$. RMSE values are reported in units of $\times 10^{-2}$.
Ratios denote DeepONet RMSE divided by CFD RMSE.}
\label{tab:rmse_summary}
\begin{tabular}{c|ccc|ccc}
\hline
$N_{\mathrm{obs}}$
& \multicolumn{3}{c|}{$A(x)$}
& \multicolumn{3}{c}{$M(x)$} \\
& CFD & DeepONet & RMSE$_A$ ratio
& CFD & DeepONet & RMSE$_M$ ratio\\
\hline
5   & 6.16 & 7.75 & 1.26
    & 5.39 & 11.4 & 2.11 \\
10  & 2.00 & 3.11 & 1.56
    & 7.75 & 5.63 & 0.73 \\
20  & 3.11 & 2.60 & 0.83
    & 4.62 & 2.00 & 0.43 \\
40  & 2.94 & 1.94 & 0.66
    & 1.18 & 2.14 & 1.81 \\
80  & 1.12 & 1.13 & 1.01
    & 0.98 & 2.11 & 2.16 \\
100 & 0.86 & 0.98 & 1.14
    & 0.71 & 1.68 & 2.36 \\
\hline
\end{tabular}
\end{table}

\subsection{Computational Efficiency and Speedup}
\label{sec:results_speedup}

The computational efficiency of the surrogate-accelerated Bayesian inference framework is assessed
by comparing the cost of likelihood evaluations using the CFD-based forward solver and the
DeepONet surrogate under identical NUTS sampling configurations. All experiments were performed on
a Linux workstation equipped with 32 CPU cores and 67~GB of RAM using CPU-only execution
(Python~3.9.23, JAX~0.4.30, NumPyro~0.19.0). For both approaches, JAX was explicitly configured to use
the CPU backend to ensure a fair and consistent comparison.

Both inference pipelines employed identical NUTS settings, consisting of 100 warm-up iterations
followed by 300 retained posterior samples using a single Markov chain. Consequently, differences
in wall-clock time directly reflect the cost of evaluating the forward model within the likelihood,
rather than differences in sampling strategy or convergence behavior.

\begin{figure}[H]
\centering
\includegraphics[width=0.85\linewidth]{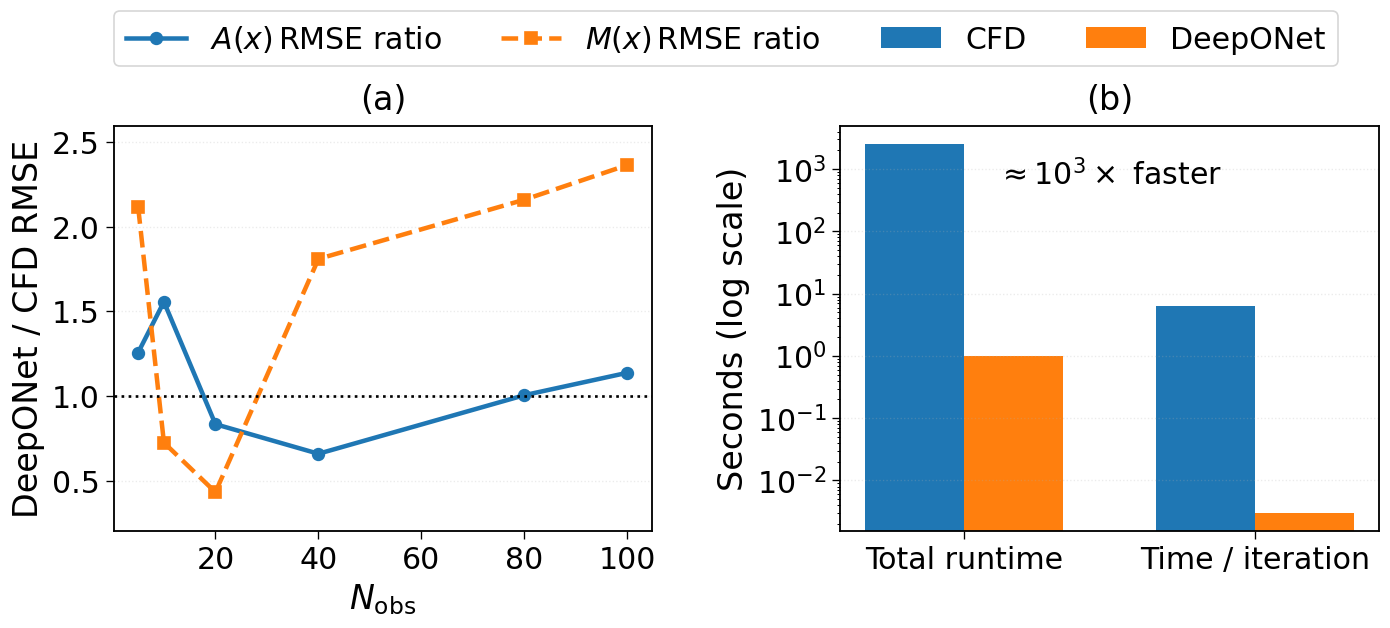}
\caption{Computational efficiency and accuracy trade-offs of surrogate-accelerated Bayesian inference.
(a) Ratio of DeepONet-based RMSE to CFD-based RMSE for the reconstructed geometry $A(x)$ and Mach field
$M(x)$ as functions of observation density $N_{\mathrm{obs}}$. The dashed horizontal line denotes parity
with the CFD-based inference. (b) Comparison of total runtime per inference run and average time per
NUTS iteration using CFD and DeepONet forward models, shown on a logarithmic scale. The surrogate-based
approach achieves an acceleration of approximately three orders of magnitude in both metrics.}
\label{fig:speedup_summary}
\end{figure}

Fig.~\ref{fig:speedup_summary}(b) reports the total runtime and average time per NUTS iteration for
the two approaches on a logarithmic scale. The CFD-based inference required approximately
41--43 minutes per run, corresponding to an average cost of $\sim$6.4~s per NUTS iteration. This
runtime was largely insensitive to the number of observations, since each likelihood evaluation
involved a full quasi-one-dimensional CFD solve.

In contrast, replacing the CFD solver with the DeepONet surrogate reduced the total inference time
to under one second across all tested cases, with an average cost of $\sim$3~ms per iteration.
This corresponds to a speedup of approximately three orders of magnitude in both total runtime and
per-iteration cost. The substantial acceleration arises from replacing the time-marching CFD solver
with a single neural-operator evaluation during each likelihood call.

To contextualize this acceleration with respect to inference accuracy, Fig.~\ref{fig:speedup_summary}(a)
reports the ratio of DeepONet-based RMSE to CFD-based RMSE for both the reconstructed geometry $A(x)$
and Mach field $M(x)$ as functions of the observation density $N_{\mathrm{obs}}$. Across all cases,
the surrogate-based inference maintains reconstruction errors comparable to the CFD-based reference,
with deviations primarily observed in data-sparse regimes and near shock-sensitive regions. These
differences remain modest relative to the achieved computational gains.

A numerical summary of the average computational cost and corresponding speedup is provided in
Table~\ref{tab:speedup_summary}. Overall, the results demonstrate that the DeepONet surrogate
enables Bayesian inference with near-CFD accuracy at a computational cost reduced by more than three
orders of magnitude, making high-fidelity uncertainty quantification feasible in regimes where
CFD-based inference would otherwise be prohibitively expensive.

\begin{table}[t]
\centering
\caption{Average computational cost of Bayesian inference using CFD and DeepONet forward models,
along with the resulting surrogate speedup. All timings correspond to CPU-only execution with
identical NUTS sampling configurations.}
\label{tab:speedup_summary}
\begin{tabular}{lccc}
\hline
Forward model
& Total runtime per run
& Avg. time per iteration
& Speedup \\ \hline
CFD-based solver
& $\sim$42 min
& $\sim$6.4 s / it
& -- \\
DeepONet surrogate
& $< 1$ s
& $\sim$0.003 s / it
& $\mathcal{O}(10^{3})$ \\ \hline
\end{tabular}
\end{table}

\FloatBarrier

\subsection{Direct Inverse Neural Operator: Deterministic Reconstruction}
\label{sec:results_direct_inverse}
As an alternative to Bayesian inverse inference, we consider a deterministic reconstruction strategy
based on a direct inverse neural operator that maps sparse Mach observations directly to the nozzle
geometry. Once trained, the inverse operator produces a single geometry prediction through a single
forward evaluation, without likelihood formulation, posterior sampling, or uncertainty propagation.
The results presented here are intended to illustrate how such a deterministic inverse-design
approach behaves and to assess its qualitative performance as an alternative tool for inverse design.

\begin{figure}[H]
  \centering
  \includegraphics[width=\linewidth]{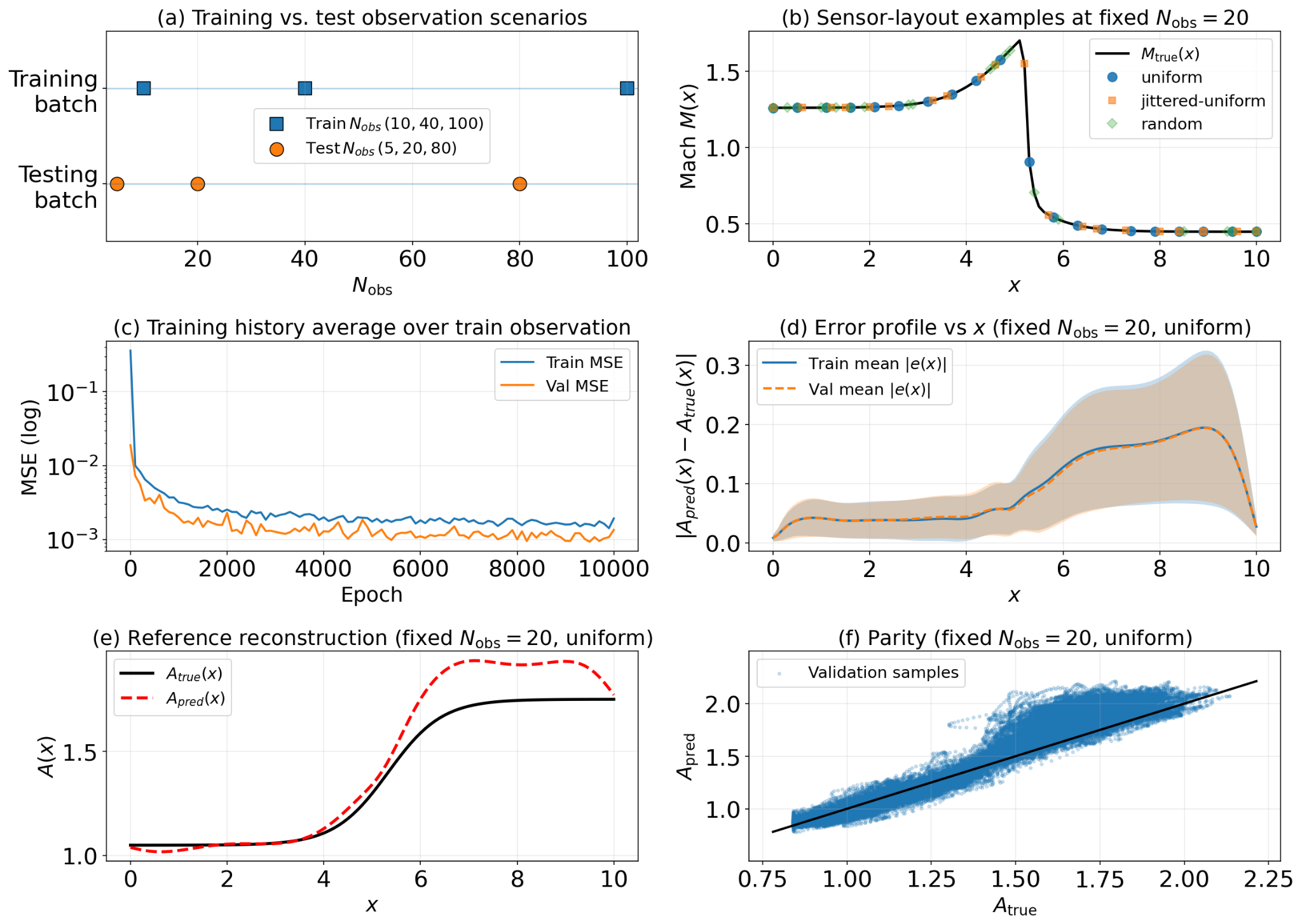}
  \caption{
  Training behavior and reconstruction characteristics of the direct inverse neural operator.
  (a) Training and testing observation densities.
  (b) Representative sensor layouts at fixed $N_{\mathrm{obs}}=20$.
  (c) Training and validation mean-squared error histories.
  (d) Mean absolute reconstruction error as a function of axial location.
  (e) Example deterministic geometry reconstruction.
  (f) Parity plot comparing predicted and true geometry values.
  }
  \label{fig:R9}
\end{figure}

\begin{figure}[H]
  \centering
  \includegraphics[width=\linewidth]{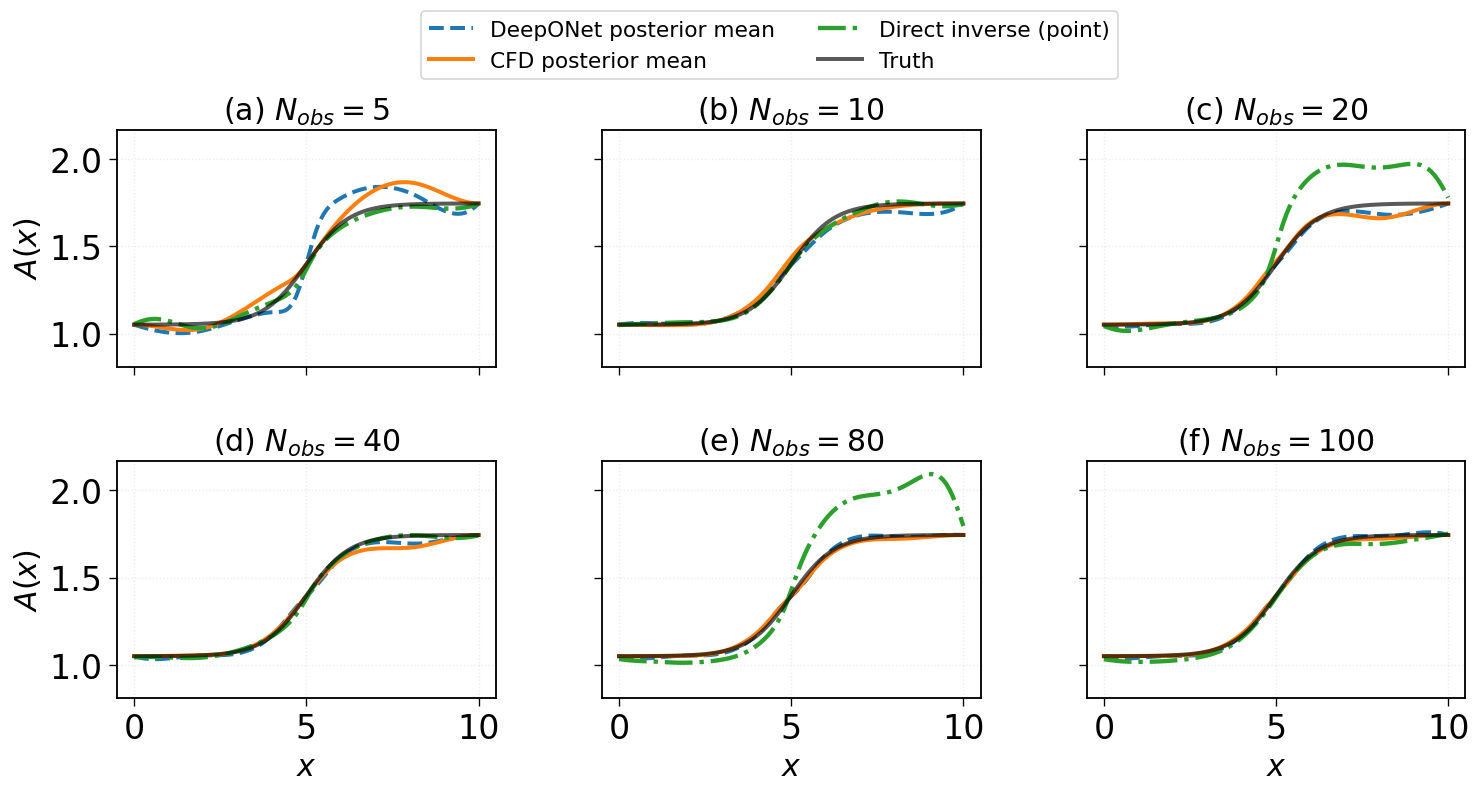}
  \caption{
  Comparison of deterministic and Bayesian nozzle geometry reconstructions for increasing observation
  density. Each panel shows the direct inverse DeepONet prediction, the DeepONet-accelerated Bayesian
  posterior mean, the CFD-based Bayesian posterior mean, and the true geometry for
  $N_{\mathrm{obs}}=5,10,20,40,80,$ and $100$.
  }
  \label{fig:R10}
\end{figure}

Figures~\ref{fig:R9} and~\ref{fig:R10} summarize the training behavior and the reconstruction of performance of the direct inverse neural operator for
geometry reconstruction from sparse Mach observations. As shown in Fig.~\ref{fig:R9}(a), the model is trained using
observation densities $N_{\mathrm{obs}}={10,40,100}$ and evaluated on previously unseen densities
$N_{\mathrm{obs}}={5,20,80}$ to assess generalization across sensor sparsity. Representative sensor
layouts for a fixed observation density are illustrated in Fig.~\ref{fig:R9}(b), including uniform,
jittered-uniform, and random placements, demonstrating the variability in spatial sampling used as
inputs to the inverse operator.

The training and validation loss histories in Fig.~\ref{fig:R9}(c) show smooth and consistent
convergence, with no visible divergence between the two curves. The spatial distribution of reconstruction error in Fig.~\ref{fig:R9}(d) indicates that the largest errors
are localized in the rapid area-transition region around $x \approx 4.5$--$5.5$ and in the downstream
shock-affected region around $x \approx 5.5$--$6.5$, where the geometry--flow coupling is strongest and
small geometric perturbations induce large changes in the Mach field. A representative deterministic
reconstruction for the unseen case $N_{\mathrm{obs}}=20$ is shown in Fig.~\ref{fig:R9}(e), where the
predicted geometry captures the overall trend of the reference solution. The
parity plot in Fig.~\ref{fig:R9}(f) shows strong correlation between predicted and true geometry values,
with increased scatter corresponding to regions of higher geometric sensitivity.

Fig.~\ref{fig:R10} compares deterministic reconstructions with Bayesian posterior means obtained
using both CFD-based and surrogate-accelerated inference across increasing observation density. For
the sparse unseen case $N_{\mathrm{obs}}=5$, the direct inverse reconstruction captures the dominant
geometric features despite limited observational information. For the moderate unseen case
$N_{\mathrm{obs}}=20$, the deterministic reconstruction remains visually consistent with the Bayesian
posterior means over most of the domain. For the dense unseen case $N_{\mathrm{obs}}=80$, localized
deviations appear in the deterministic reconstruction, particularly in the downstream region, while
the Bayesian posterior means remain smoother. For observation densities used during training
($N_{\mathrm{obs}}=10, 40$ and $100$), close agreement between deterministic and Bayesian reconstructions is
observed.

Overall, these figures illustrate how a direct inverse neural operator can serve as a fast,
deterministic alternative for inverse design, while also highlighting differences relative to
Bayesian approaches that explicitly account for uncertainty.

\FloatBarrier

\FloatBarrier

\section{Conclusion}
\label{sec:conclusions}

This work demonstrates that neural operator surrogates can substantially reduce the computational
burden associated with fully Bayesian inverse design for PDE-constrained flow problems.
By embedding a DeepONet-based neural operator directly within a gradient-based Markov Chain Monte
Carlo framework, we enable uncertainty-aware inverse nozzle design at a computational cost reduced
by more than three orders of magnitude relative to a CFD-based reference, while preserving the
essential structure of the Bayesian posterior.

A CFD-based Bayesian inverse formulation was first established for quasi--one-dimensional nozzle flow,
providing a high-fidelity reference for geometry reconstruction under uncertainty.
The effect of geometry parameterization was examined, and among the discrete, polynomial, and cubic
B-spline representations considered, the cubic B-spline was shown to yield the most stable posterior
behavior and lowest reconstruction error. This representation was therefore adopted for all
subsequent surrogate training and inference.

A DeepONet surrogate trained on CFD-generated data was then substituted for the CFD solver within the
Bayesian likelihood, without altering the probabilistic formulation or sampling strategy.
Across sparse to fully observed regimes, the surrogate-accelerated inference accurately reproduced
the posterior mean geometry and uncertainty trends obtained using the CFD solver.
Quantitative error metrics confirmed close agreement between CFD-based and DeepONet-based inference,
with modest discrepancies confined to regions of strong nonlinearity.
Most importantly, the computational cost of Bayesian inference was reduced from tens of minutes to
sub-second execution on a CPU-only platform, demonstrating the practicality of neural
operator--accelerated Bayesian inversion.

In addition to probabilistic inference, a direct inverse neural operator was explored as a
deterministic alternative for inverse design.
Although this approach does not provide uncertainty quantification, the results illustrate that
single-shot inverse prediction can recover meaningful geometric features across a range of
observation densities, including cases not encountered during training.
These findings position direct inverse operators as a complementary tool for rapid inverse design,
while highlighting the advantages of Bayesian formulations when uncertainty information is required.

Several limitations remain.
The surrogate models are trained on a prescribed distribution of geometries and therefore exhibit
their highest fidelity within that distribution.
Approximation errors become more apparent as the posterior concentrates in strongly nonlinear
regions, such as near shock locations.
Moreover, the present study is limited to steady, quasi--one-dimensional flow and does not address
time-dependent effects or higher-dimensional configurations.

Future work will extend this framework to multidimensional and unsteady flows, investigate
multi-fidelity and adaptive surrogate strategies, and incorporate physics-informed or
structure-preserving neural operator architectures. Extensions to surrogate frameworks, such as basis-to-basis operator models, may further improve surrogate fidelity and generalization for strongly nonlinear inverse problems~\cite{ingebrand2025basis}.
Such extensions will further improve robustness and generalization, enabling scalable,
uncertainty-aware inverse design for increasingly complex aerodynamic systems.

Overall, this study establishes neural operator--accelerated Bayesian inference as a practical and
scalable approach for inverse design under uncertainty, bridging the gap between high-fidelity CFD
and modern probabilistic design methodologies.

\section*{Funding Declaration}
This work was supported in part by the Air Force Office of Scientific Research 
(AFOSR) under Grant No.\ FA9550\textendash 24\textendash 1\textendash 0327. 


\section*{Author Contributions}

O.\,S.\ conceptualized the study, provided theoretical guidance, and supervised the project. B.\,T.\ designed and executed the neural operator framework, performed the numerical experiments, carried out the analyses, and prepared the initial draft. Both authors discussed the results, contributed to the writing, and approved the final manuscript.


\section*{Competing Interests}

The authors declare no competing interests.

\section*{Data Availibility}

All codes, experimental findings, and trained model results associated with this work are publicly available in our GitHub repository: 
\url{https://github.com/bipintiwari2950/Neural-Operator---Bayesian-Inverse-CFD}.






\begin{appendices}

\section{Forward Solver For CFD}\label{secA1}

All governing equations and solver variables in this appendix are written in nondimensional form using the scaling defined in the Forward Operator methodology section. This appendix provides the full mathematical and numerical formulation of the forward solver used to evaluate the mapping
\begin{equation}
F: A(x) \longrightarrow M(x),
\label{eq:F_mapping}
\end{equation}
which computes the Mach number distribution corresponding to a given nozzle area profile.

\subsection{Governing Equations}

The quasi--1D Euler equations for compressible flow in a variable-area duct are:

\begin{equation}
\frac{\partial (\rho A)}{\partial t} + 
\frac{\partial (\rho u A)}{\partial x} = 0,
\label{eq:mass}
\end{equation}

\begin{equation}
\frac{\partial (\rho u A)}{\partial t}
+ \frac{\partial \left( A(\rho u^2 + P)\right)}{\partial x}
= P \frac{dA}{dx},
\label{eq:momentum}
\end{equation}

\begin{equation}
\frac{\partial (E A)}{\partial t}
+ \frac{\partial \left( u A (E+P) \right)}{\partial x}
= 0.
\label{eq:energy}
\end{equation}
where $\rho$ is the density, $u$ is the velocity, $P$ is the pressure, $E$ is the total energy per unit volume, and $A(x)$ is the nozzle cross-sectional area. The total energy is
\begin{equation}
E = \frac{P}{\gamma - 1} + \frac{1}{2} \rho u^2,
\end{equation}
with $\gamma$ the ratio of specific heats.

\subsection{Nozzle Geometry}
The nozzle area $A(x)$ and its derivative are defined as
\begin{align}
A(x) &= 1.398 + 0.347 \tanh(0.8 x - 4), \\
\frac{dA}{dx} &= 0.2766 \left( 1 - \tanh^2(0.8 x - 4) \right).
\end{align}

\subsection{Numerical Method}
\subsubsection{Discretization}
The domain is discretized into $N$ control volumes with cell centers $x_i$ and spacing
\begin{equation}
\Delta x_i = x_{i+1} - x_i, \quad \Delta x_N = \Delta x_{N-1}.
\end{equation}

\subsubsection{Conservative and Primitive Variables}
Define the conservative vector:
\begin{equation}
\mathbf{Q}_i = 
\begin{bmatrix}
\rho_i A_i \\ \rho_i u_i A_i \\ E_i A_i
\end{bmatrix},
\end{equation}
and the primitive variables are recovered as
\begin{align}
\rho_i &= \frac{Q_{i,1}}{A_i}, &
u_i &= \frac{Q_{i,2}}{Q_{i,1}}, &
E_i &= \frac{Q_{i,3}}{A_i}, &
P_i &= (\gamma - 1)\left(E_i - \frac{1}{2} \rho_i u_i^2\right).
\end{align}

\subsubsection{Flux Splitting}
The upwind flux splitting uses Mach-number dependent splitting to handle subsonic and supersonic flows:
\begin{equation}
M_i = \frac{u_i}{c_i}, \quad c_i = \sqrt{\frac{\gamma P_i}{\rho_i}}
\end{equation}
where $c_i$ is the local sound speed. The pressure and momentum fluxes are split as:

\[
P_i^+ =
\begin{cases}
\frac{1}{2} (1 + M_i) P_i, & |M_i| < 1 \\
P_i, & M_i > 1 \\
0, & M_i < -1
\end{cases}, \quad
P_i^- =
\begin{cases}
\frac{1}{2} (1 - M_i) P_i, & |M_i| < 1 \\
0, & M_i > 1 \\
P_i, & M_i < -1
\end{cases}
\]

\[
(u P)_i^+ =
\begin{cases}
\frac{1}{2} (u_i + c_i) P_i, & |M_i| < 1 \\
u_i P_i, & M_i > 1 \\
0, & M_i < -1
\end{cases}, \quad
(u P)_i^- =
\begin{cases}
\frac{1}{2} (u_i - c_i) P_i, & |M_i| < 1 \\
0, & M_i > 1 \\
u_i P_i, & M_i < -1
\end{cases}
\]
The upwind fluxes are then
\begin{align}
F_i^+ &= 
\begin{bmatrix}
\rho_i u_i A_i \\
(\rho_i u_i^2 + P_i^+) A_i \\
(u_i E_i + (u P)_i^+) A_i
\end{bmatrix}, &
F_i^- &=
\begin{bmatrix}
\rho_i u_i A_i \\
(\rho_i u_i^2 + P_i^-) A_i \\
(u_i E_i + (u P)_i^-) A_i
\end{bmatrix},
\end{align}
with source term
\begin{equation}
H_i =
\begin{bmatrix} 0 \\ P_i \frac{dA}{dx}_i \\ 0 \end{bmatrix}.
\end{equation}

\subsubsection{Time-stepping and CFL Condition}
The timestep is computed from the CFL condition:
\begin{equation}
\Delta t = \text{CFL} \times \min_i \frac{\Delta x_i}{|u_i| + c_i}.
\end{equation}

\subsubsection{Update Equation}
The conservative variables are updated as
\begin{equation}
\mathbf{Q}_i^{n+1} = \mathbf{Q}_i^n - \frac{\Delta t}{\Delta x_i} \left(F_i^+ - F_{i-1}^+ + F_{i+1}^- - F_i^- \right) + \Delta t H_i.
\end{equation}

\subsection{Boundary Conditions}

\paragraph{Supersonic inlet:}
All primitive variables $\rho_0, u_0, P_0$ are prescribed.

\paragraph{Subsonic outlet:}
\begin{equation}
P_N = P_{\text{out}},
\label{eq:outlet_pressure}
\end{equation}

\begin{equation}
\rho_N = 2\rho_{N-1} - \rho_{N-2},
\qquad
u_N   = 2u_{N-1} - u_{N-2}.
\label{eq:outlet_extrap}
\end{equation}

\subsection{Algorithmic Summary and Solver Outputs}

The forward solver computes steady-state solutions of the quasi--one-dimensional Euler equations by iteratively updating the conservative variables using the finite-volume formulation. Starting from an initial guess for the primitive variables $(\rho, u, P)$, the solver converts them into conservative form, evaluates the speed of sound, Mach number, geometric source terms, and Mach-split numerical fluxes,. The timestep is selected using the CFL stability condition, after which the conservative variables are advanced using finite-volume update equation as described above. Subsequently, the primary variables are recovered and boundary conditions are imposed at the inlet and outlet. Convergence to steady state is assessed by monitoring the normalized residual norm until a specified tolerance is reached.

\begin{figure}[H]
    \centering
    \includegraphics[width=1\textwidth]{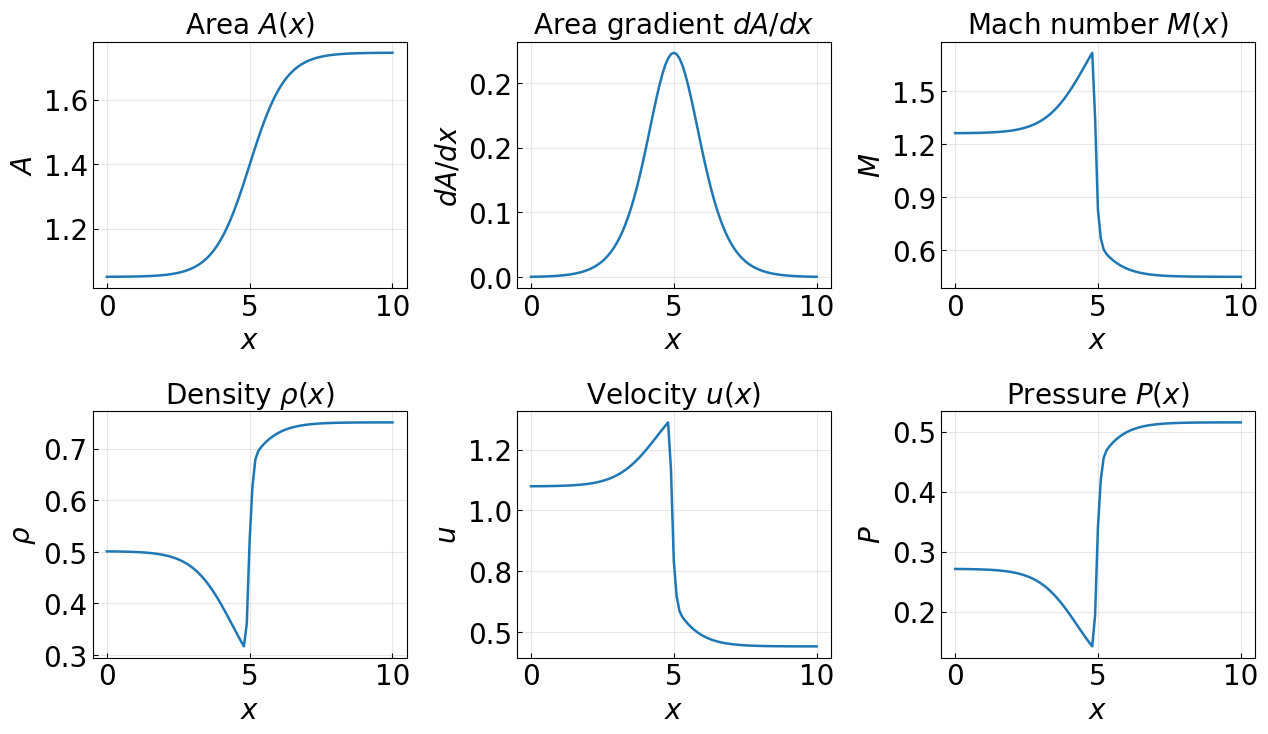}
    \caption{Representative forward-solver outputs for a smooth nozzle geometry. 
    Panels show the area $A(x)$, density $\rho(x)$, velocity $u(x)$, pressure $P(x)$, and Mach number $M(x)$.}
    \label{fig:solver_outputs_example}
\end{figure}

The iterative procedure yields steady-state distributions of density $\rho(x)$, velocity $u(x)$, pressure $P(x)$ and Mach number $M(x)$ with the Mach profile serving as the primary quantity used in inverse reconstruction and surrogate modeling. Fig.~\ref{fig:solver_outputs_example} provides a representative example showing the distribution of the nozzle area and the corresponding flow fields produced by the solver.

\FloatBarrier

\subsection{\BT{Exact Analytical Solution and Forward Solver Validation}}
\BT{To rigorously validate the numerical fidelity of the finite-volume quasi-one-dimensional Euler solver described above, its predictions are compared against the exact analytical solution for compressible nozzle flow with a standing normal shock. Unlike the numerical solver, which employs flux-splitting and introduces artificial dissipation to stably capture discontinuities over several grid cells, the analytical solution treats the flow as perfectly isentropic everywhere except at the shock, which is modeled as an infinitely thin mathematical discontinuity.}

\BT{The exact solution is constructed by matching two isentropic flow branches across a normal shock. For any axial location $x$ upstream or downstream of the shock, steady, inviscid, and isentropic compressible flow dictates that the local Mach number $M(x)$ and nozzle area $A(x)$ satisfy the area--Mach relation:}
\begin{equation}
\BT{\frac{A(x)}{A^*} =
\frac{1}{M(x)}
\left[
\frac{2}{\gamma+1}
\left( 1 + \frac{\gamma-1}{2}M(x)^2 \right)
\right]^{\frac{\gamma+1}{2(\gamma-1)}}}
\label{eq:area_mach}
\end{equation}
\BT{where $A^*$ is the reference sonic throat area and $\gamma$ is the ratio of specific heats.}

\BT{Upstream of the shock, the flow accelerates supersonically governed by the upstream sonic area $A_1^*$, determined by the prescribed inlet conditions. At an arbitrary shock location $x_s$, the flow undergoes a discontinuous transition. The Mach number immediately downstream of the shock, $M_2$, is related to the upstream Mach number, $M_1$, by the normal shock relation:}
\begin{equation}
\BT{M_2^2 =
\frac{1 + \frac{\gamma-1}{2} M_1^2}
{\gamma M_1^2 - \frac{\gamma-1}{2}}}
\end{equation}

\BT{This non-isentropic transition results in a loss of stagnation pressure ($P_{02} < P_{01}$) and a proportional increase in the required sonic throat area ($A_2^* > A_1^*$). The stagnation pressure ratio across the shock is given by}
\begin{equation}
\BT{\frac{P_{02}}{P_{01}} =
\left(
\frac{(\gamma+1)M_1^2}{(\gamma-1)M_1^2+2}
\right)^{\frac{\gamma}{\gamma-1}}
\left(
\frac{\gamma+1}{2\gamma M_1^2-(\gamma-1)}
\right)^{\frac{1}{\gamma-1}}}
\end{equation}
\BT{which dictates the downstream sonic area $A_2^* = A_1^*(P_{01}/P_{02})$.}

\BT{Downstream of the shock ($x > x_s$), the flow is again isentropic, governed by Equation~\ref{eq:area_mach} using $A_2^*$. The exact physical location of the normal shock $x_s$ is determined by iterating until the exit pressure $P(L)$ perfectly matches the prescribed subsonic outlet boundary condition $P_{\text{out}}$.}

\BT{Once $M(x)$ and the regional stagnation pressure ($P_{01}$ upstream or $P_{02}$ downstream) are established, the local static pressure $P(x)$ and temperature $T(x)$ are recovered via isentropic relations:}
\begin{equation}
\BT{P(x) =
P_0
\left(
1 + \frac{\gamma-1}{2} M(x)^2
\right)^{-\frac{\gamma}{\gamma-1}}}
\end{equation}

\begin{equation}
\BT{T(x) =
T_0
\left(
1 + \frac{\gamma-1}{2} M(x)^2
\right)^{-1}}
\end{equation}

\BT{Because the flow is adiabatic, the stagnation temperature $T_0$ remains constant across the entire domain. The local static temperature $T(x)$ allows the exact local density $\rho(x)$ and velocity $u(x)$ to be computed directly using the ideal gas law and the definition of the Mach number:}
\begin{equation}
\BT{\rho(x) = \frac{P(x)}{T(x)},
\qquad
u(x) = M(x)\sqrt{\gamma T(x)}}
\end{equation}

\begin{figure}[H]
    \centering
    \includegraphics[width=0.75\textwidth]{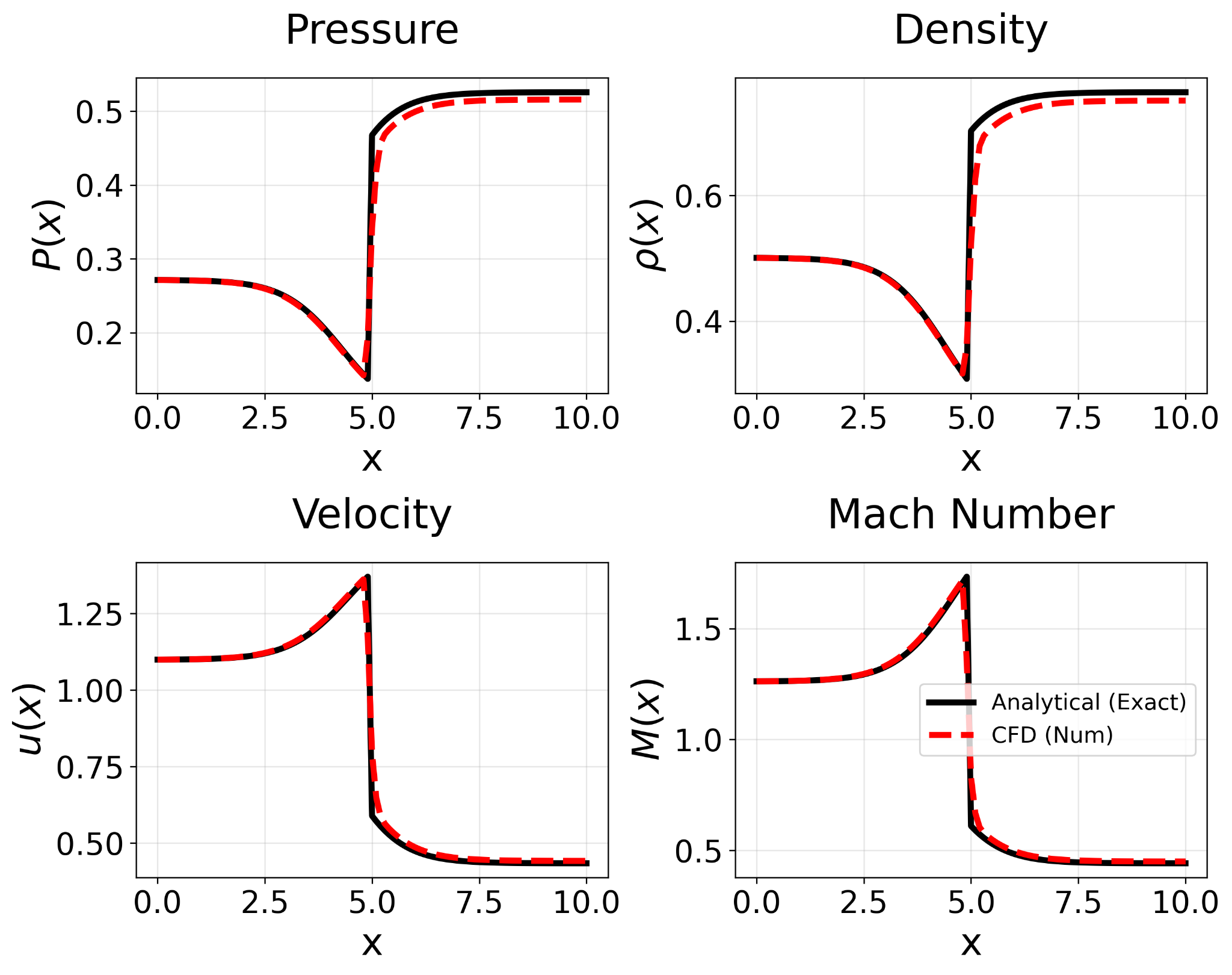} 
    \caption{\BT{Steady-state validation of the JAX-based finite-volume CFD solver against the exact analytical solution. The panels compare the spatial profiles of pressure, density, velocity, and Mach number. The numerical predictions accurately capture the smooth isentropic acceleration and deceleration branches. The observed smearing at the shock location ($x = 5.0$) and the minor downstream offsets are expected physical artifacts of the first-order upwind flux-splitting scheme, which introduces necessary numerical dissipation to stably capture the discontinuity without spurious oscillations.}}
    \label{fig:solver_validation_exact}
\end{figure}

\BT{A comprehensive derivation of these fundamental compressible flow relations can be found in the literature \cite{hoffmann2000computational, anderson2003modern}. As shown in Fig.~\ref{fig:solver_validation_exact}, the numerical solver accurately captures both flow branches. The absolute error is effectively zero in smooth, isentropic regions, with observable deviations confined strictly to the shock location due to the first-order upwind flux splitting scheme smearing gradients to maintain numerical stability. This confirms that the forward operator provides a physically consistent and reliable mapping $A(x) \mapsto M(x)$ for the inverse design framework.}

\FloatBarrier

\section{Geometry Parameterizations}
\label{app:geom_param}

This appendix describes the geometric parameterizations used to represent the distribution of the nozzle area $A(x)$ in both forward simulations and surrogate model training. All representations satisfy exact endpoint value and slope constraints and are defined on the nondimensional axial coordinate $x$. The nondimensional axial coordinate $x$ spans a fixed computational domain from $x = 0$ at the inlet to $x = L$ at the outlet, where the total length of the nondimensional nozzle is L = 10. This is consistent with the normalization used throughout the main text. The three parameterizations differ in smoothness, locality, and numerical conditioning, enabling a systematic assessment of their impact on inverse estimation and surrogate model performance.

\subsection{Discrete Piecewise-Linear Representation}

In the discrete representation, the nozzle area is specified at a finite set of axial locations \(\{x_j\}_{j=0}^{K-1}\) with corresponding values \(\{A_j\}\). Between adjacent nodes, the area is interpolated linearly as
\begin{equation}
A(x) = A_j + \frac{A_{j+1}-A_j}{x_{j+1}-x_j}(x-x_j),
\qquad x_j \le x \le x_{j+1},
\label{eq:discA}
\end{equation}
with piecewise-constant derivative
\begin{equation}
A'(x) = \frac{A_{j+1}-A_j}{x_{j+1}-x_j}.
\label{eq:discAprime}
\end{equation}

\BT{In this representation, the geometry parameter vector introduced in Section~\ref{sec:geometry_parameterization} corresponds to the nodal area values defining the nozzle geometry. Specifically, the parameter vector is given by
\begin{equation}
\boldsymbol{\theta} = (A_1, A_2, \ldots, A_{K-2}),
\end{equation}
while the endpoint values remain fixed by the imposed boundary conditions.} This representation is $C^0$ continuous and strictly local: each parameter influences the geometry only within a single interval. While this provides maximal flexibility, the lack of smoothness can lead to numerical stiffness and increased sensitivity to noise in inverse problems and gradient-based optimization. 
\subsection{Polynomial Representation}

A globally smooth representation is obtained by expressing the nozzle area as a seventh-degree polynomial in the nondimensional axial coordinate $x$,
\begin{equation}
A(x) = \sum_{k=0}^{7} b_k x^k,
\qquad
A'(x) = \sum_{k=1}^{7} k\,b_k\,x^{k-1}.
\label{eq:polyAcompact}
\end{equation}

Boundary conditions on the inlet and outlet area and slope impose the constraints
\begin{equation}
\begin{aligned}
b_0 &= A_0, &
b_1 &= A'_0, \\
\sum_{k=0}^{7} b_k\,L^k &= A_L, &
\sum_{k=1}^{7} k\,b_k\,L^{k-1} &= A'_L,
\end{aligned}
\label{eq:poly_constraints}
\end{equation}
which uniquely determine the highest-order coefficients,
\begin{equation}
\begin{aligned}
b_6 &= -A'_L -6b_1 -5b_2 -4b_3 -3b_4 -2b_5 + 7(A_L-A_0), \\
b_7 &= A_L - \sum_{k=0}^{6} b_k\,L^k.
\end{aligned}
\label{eq:poly_b6b7}
\end{equation}

The remaining coefficients $b_2, b_3, b_4, b_5$ serve as free parameters. \BT{Accordingly, the geometry parameter vector introduced in Section~\ref{sec:geometry_parameterization} corresponds to these polynomial coefficients, i.e.,
\begin{equation}
\boldsymbol{\theta} = (b_2, b_3, b_4, b_5),
\end{equation}
which uniquely determine the nozzle area distribution once the endpoint constraints are enforced.} This representation yields a globally smooth geometry but exhibits global support, meaning that perturbations to any coefficient affect the entire nozzle. Such nonlocal coupling can degrade numerical conditioning and interpretability in inverse problems.

\subsection{Cubic B-Spline Representation}
The cubic B-spline representation combines global smoothness with strong local control. 
The distribution of the nozzle area is expressed as
\begin{equation}
A(x) = \sum_{i=0}^{K-1} N_{i,3}(x;\boldsymbol{\tau})\,c_i,
\qquad
A'(x) = \sum_{i=0}^{K-1} N'_{i,3}(x;\boldsymbol{\tau})\,c_i,
\label{eq:bsAcompact}
\end{equation}
where $c_i$ are the $K$ spline control coefficients and $N_{i,3}$ are cubic (degree--3) B-spline basis functions. 
The basis functions are defined on the knot sequence
\begin{equation}
\boldsymbol{\tau} = \{\tau_0,\tau_1,\ldots,\tau_m\},
\qquad
m = K + 3,
\end{equation}
so that the total number of knots is $m+1 = K+4$. 
The knot sequence partitions the spatial domain $[0,L]$ and determines the compact support of each basis function, ensuring that each coefficient $c_i$ influences the geometry only over a limited axial region.

A clamped (open) knot sequence is employed, in which the first and last knots are repeated with multiplicity $p+1=4$. 
This construction enforces interpolation at the domain boundaries and guarantees exact satisfaction of endpoint constraints,
\begin{equation}
A(0) = c_0 = A_0,
\qquad
A(L) = c_{K-1} = A_L .
\label{eq:bs_endpoints}
\end{equation}

For cubic splines, the endpoint slopes depend only on adjacent control points and the local knot spacing,
\begin{equation}
A'(0) = \frac{3}{\tau_4 - \tau_1}\left(c_1 - c_0\right) = A'_0,
\qquad
A'(L) = \frac{3}{\tau_m - \tau_{m-3}}\left(c_{K-1} - c_{K-2}\right) = A'_L .
\label{eq:bs_slopes}
\end{equation}
These relations yield the constrained coefficients
\begin{equation}
c_1 = A_0 + \frac{\tau_4 - \tau_1}{3} A'_0,
\qquad
c_{K-2} = A_L - \frac{\tau_m - \tau_{m-3}}{3} A'_L .
\label{eq:bs_fixed}
\end{equation}

After enforcing endpoint value and slope constraints, the remaining interior coefficients are parameterized as
\begin{equation}
c_i = c_i^{\mathrm{base}} + \delta_i,
\qquad
2 \le i \le K-3,
\label{eq:bs_free}
\end{equation}
where $c_i^{\mathrm{base}}$ defines a smooth reference geometry and $\delta_i$ represent localized perturbations. \BT{In this representation, the geometry parameter vector introduced in Section~\ref{sec:geometry_parameterization} corresponds to the spline perturbation coefficients, i.e.,
\begin{equation}
\boldsymbol{\theta} = (\delta_2, \delta_3, \ldots, \delta_{K-3}),
\end{equation}
which determine the interior control points of the cubic B-spline representation and therefore define the nozzle geometry.} 
This formulation yields a $C^2$-continuous nozzle geometry with compact support and well-conditioned local sensitivity, 
making the cubic B-spline representation particularly suitable for ill-posed inverse problems CFD or operator learning-based surrogate modeling.

\subsection{Summary}

The discrete representation provides maximum locality without smoothness. The seventh-degree polynomial representation offers global smoothness but introduces strong global coupling, in the sense that perturbations to any coefficient affect the geometry over the entire domain. In contrast, the cubic B-spline representation achieves smoothness and locality simultaneously through compactly supported basis functions. These structural differences have direct implications for numerical conditioning, identifiability, and the performance of data-driven inverse problems.


\section{DeepONet Architecture and Training Details}
\label{sec:deeponet_appendix}

This appendix describes the operator-learning methodology used for both the forward and inverse neural
operators employed in this work. DeepONets are used to approximate
the nonlinear forward mapping $\widehat{F}: A(x)\mapsto M(x)$ as well as a deterministic inverse mapping
$\widehat{G}: (M_{\mathrm{sparse}}(x),\omega(x))\mapsto A(x)$. The governing equations, numerical
discretization, and boundary conditions of the CFD solver are provided in
Appendix~\ref{secA1}.

Both neural operators are trained using CFD-labeled geometry--flow pairs
$\{A^{(j)}(x),M^{(j)}(x)\}$ generated via the cubic B-spline parameterization described in
Appendix~\ref{app:geom_param}.

\subsection{Dataset Generation via B-Spline Sampling and CFD Evaluation}
\label{sec:deeponet_data}

All nozzle geometries are discretized on the same axial grid \(\{x_i\}_{i=1}^{N}\) spanning $x \in [0,10]$, identical to the grid used by the CFD solver
(Appendix~\ref{secA1}). Therefore, each geometry is represented as a vector of
\(\mathbf{A} = [A(x_1),\dots,A(x_N)]^\top \in \mathbb{R}^{N}\)
and a corresponding Mach profile
\(\mathbf{M} = [M(x_1),\dots,M(x_N)]^\top \in \mathbb{R}^{N}\) with \(N=101\) points for all samples.

Training and validation geometries are generated using the cubic B-spline representation
(Eqs.~\eqref{eq:bsAcompact}--\eqref{eq:bs_free}) with \(K=10\) control coefficients and a clamped knot
sequence enforcing exact interpolation at \(x=0\) and \(x=L\) (Eq.~\eqref{eq:bs_endpoints}). To improve geometric resolution in dynamically sensitive regions, interior knots are nonuniformly distributed and clustered around the $x=5.0$ with a clustering width of 1.5 to enhance geometric resolution in regions of strong flow sensitivity. A smooth baseline control vector \(c^{\mathrm{base}}\) is constructed between the inlet and outlet areas,
and interior coefficients are perturbed according to
\begin{equation}
c_i = c_i^{\mathrm{base}} + \delta_i, \qquad 2 \le i \le K-3,
\label{eq:deeponet_bs_sampling}
\end{equation}
where \(\delta_i\) are independent Gaussian random variables with zero mean and prescribed standard deviation $\sigma_c = 0.2$ which directly controls the magnitude of geometric variation. This sampling strategy yields smooth, physically plausible nozzle variations with localized control while preserving global geometric structure. The endpoint coefficients \(c_0\) and \(c_{K-1}\) remain fixed, ensuring consistent endpoint constraints.
Fig.~\ref{fig:appendixC_deeponet}(a--b) illustrates the control-point perturbation mechanism and representative nozzle profiles drawn directly from the training set.

Each geometry sampled \(A^{(j)}(x)\) is assigned to a Mach profile \(M^{(j)}(x)\) by running the CFD solver
(Appendix~\ref{secA1}) with identical numerical parameters (grid resolution, CFL number, iteration count 
and boundary conditions).

The final dataset contains \(n_{\mathrm{train}}=1500\) training samples and \(n_{\mathrm{val}}=500\)
validation samples, generated using the same B-spline sampling distribution but independent random seeds. The full dataset, including the reference nozzle and its CFD solution, is generated once and reused for training, validation, and evaluation, ensuring strict reproducibility of all reported results. The generated sample nozzle area is presented in Fig.~\ref{fig:appendixC_deeponet}(b).

The same CFD-generated dataset is used to train both forward and inverse DeepONet models. For the
inverse operator, sparse observation patterns are applied dynamically during training, while the
underlying geometry--flow pairs remain unchanged.
\subsection{Operator-Learning Formulation}
\label{sec:deeponet_formulation}

The DeepONet formulation described here provides a unified operator-learning framework for both the
forward and inverse neural operators used in this work. In both cases, the operator is represented as
a mapping between functions rather than finite-dimensional vectors, using a shared branch--trunk
architecture and a separable latent representation. The forward and inverse operators differ only in
(i) the choice of input function provided to the branch network and (ii) the training objective used
to fit the network parameters.

For the forward problem, we seek an approximation $\widehat{F}$ of the CFD operator $F$ such that
\begin{equation}
\widehat{M}_{\boldsymbol{\phi}}(x;A) \approx M(x) = F(A)(x),
\end{equation}
where $A(x)$ denotes the nozzle geometry and $\boldsymbol{\phi}$ are trainable network parameters.
For the inverse problem, the same operator-learning structure is employed to approximate a deterministic
mapping from sparse flow observations to geometry, as described in
Section~\ref{sec:inverse_neural_operator}.

DeepONet represents an operator through a separable latent expansion evaluated at spatial query points
$x$,
\begin{equation}
\widehat{M}_{\boldsymbol{\phi}}(x;A)
=
\sum_{k=1}^{p}
b_{\boldsymbol{\phi},k}(A)\; t_{\boldsymbol{\phi},k}(x) + \beta,
\label{eq:deeponet_core}
\end{equation}
where $\mathbf{b}_{\boldsymbol{\phi}}(A)\in\mathbb{R}^{p}$ is the branch-network embedding of the input
function and $\mathbf{t}_{\boldsymbol{\phi}}(x)\in\mathbb{R}^{p}$ is the trunk-network embedding of the
spatial coordinate. The scalar $\beta$ denotes a learned bias term.

For each input function (e.g., a geometry field for the forward operator or a sparsely observed Mach
field and mask for the inverse operator), the branch network is evaluated once to produce the latent
coefficients $\mathbf{b}_{\boldsymbol{\phi}}$. The trunk network is evaluated independently at each
spatial location $x_i$, enabling pointwise reconstruction of the output field while preserving the
operator structure.
\subsection{Network Architecture and Normalization}
\label{sec:deeponet_arch}

Both the forward and inverse neural operators differ only in the input provided to the branch network, while the trunk network
and latent dimension remain identical. For the forward operator, the branch network takes the discretized nozzle geometry
$\mathbf{A}\in\mathbb{R}^{N}$ as input, where $N=101$ denotes the number of axial grid points.
For the inverse operator, the branch network instead takes the concatenated input
$\mathbf{z}=[M_{\mathrm{sparse}}(x),\,\omega(x)]\in\mathbb{R}^{2N}$, representing the sparsely observed
Mach field and its associated observation mask. In both cases, the branch network consists of two
hidden layers with 128 neurons each and $\tanh(\cdot)$ activation functions, followed by a linear
output layer of dimension $p=64$.

The trunk network takes the scalar spatial coordinate $x\in\mathbb{R}$ as input and consists of two
hidden layers with 64 neurons each, followed by a linear output layer of the same latent dimension
$p=64$. This shared latent dimension enables the separable inner-product representation in
Eq.~\eqref{eq:deeponet_core} for both forward and inverse mappings.

Input normalization is applied to improve numerical conditioning and training stability. Geometry
inputs are normalized as
\begin{equation}
\widetilde{\mathbf{A}} = \frac{\mathbf{A}-\mu_A}{\sigma_A},
\end{equation}
where $\mu_A$ and $\sigma_A$ are computed from the union of training and validation geometries. The
spatial coordinate is normalized to the interval $[-1,1]$ via
\begin{equation}
\widetilde{x} = 2\left(\frac{x-x_{\min}}{x_{\max}-x_{\min}}-\frac{1}{2}\right),
\label{eq:deeponet_norm}
\end{equation}
where $[x_{\min},x_{\max}]$ denotes the axial domain. All normalization constants and architectural
hyperparameters are saved alongside the trained models to ensure reproducibility.

\subsection{Training Objective and Optimization}
\label{sec:deeponet_training}

The forward and inverse DeepONets are trained independently using supervised learning, each with
a mean-squared-error objective tailored to the corresponding operator.

For the forward operator, given training pairs
$\{(\mathbf{A}^{(j)},\mathbf{M}^{(j)})\}_{j=1}^{n_{\mathrm{train}}}$,
the network parameters $\boldsymbol{\phi}$ are obtained by minimizing the mean-squared error over
Mach profiles,
\begin{equation}
\mathcal{L}_{\mathrm{fwd}}(\boldsymbol{\phi})
=
\frac{1}{n_{\mathrm{train}}}
\sum_{j=1}^{n_{\mathrm{train}}}
\left(
\frac{1}{N}\sum_{i=1}^{N}
\left(
\widehat{M}_{\boldsymbol{\phi}}(x_i;\mathbf{A}^{(j)}) - M^{(j)}(x_i)
\right)^2
\right).
\label{eq:deeponet_loss_fwd}
\end{equation}

For the inverse operator, training pairs
$\{(\mathbf{z}^{(j)},\mathbf{A}^{(j)})\}$ are constructed from sparsified Mach fields and corresponding
geometries. The inverse DeepONet is trained by minimizing the reconstruction error of the geometry,
\begin{equation}
\mathcal{L}_{\mathrm{inv}}(\boldsymbol{\phi})
=
\frac{1}{n_{\mathrm{train}}}
\sum_{j=1}^{n_{\mathrm{train}}}
\left(
\frac{1}{N}\sum_{i=1}^{N}
\left(
\widehat{A}_{\boldsymbol{\phi}}(x_i;\mathbf{z}^{(j)}) - A^{(j)}(x_i)
\right)^2
\right).
\label{eq:deeponet_loss_inv}
\end{equation}

Both networks are trained using the Adam optimizer with a learning rate of $10^{-3}$. Validation loss is evaluated periodically on a held-out validation set,
using the same objective functions as in Eqs.~\eqref{eq:deeponet_loss_fwd} and
\eqref{eq:deeponet_loss_inv}. The training history of the forward and inverse DeepONet is shown in
Fig.~\ref{fig:appendixC_deeponet}(c) and Fig.~\ref{fig:R9}(c).

\subsection{Evaluation and Diagnostics}
\label{sec:deeponet_eval}

Surrogate accuracy is quantified using mean absolute error relative to the CFD reference. 
For the forward operator, the error in the Mach number at each spatial location $x_i$ is defined as
\begin{equation}
e_M(x_i)
=
\frac{1}{n_s}\sum_{j=1}^{n_s}
\left|
\widehat{M}_{\boldsymbol{\phi}}(x_i;\mathbf{A}^{(j)}) - M^{(j)}(x_i)
\right|.
\label{eq:error_profile_M}
\end{equation}
For the inverse operator, an analogous metric is used to quantify reconstruction error in the nozzle
geometry,
\begin{equation}
e_A(x_i)
=
\frac{1}{n_s}\sum_{j=1}^{n_s}
\left|
\widehat{A}_{\boldsymbol{\phi}}(x_i;\mathbf{M}^{(j)}) - A^{(j)}(x_i)
\right|.
\label{eq:error_profile_A}
\end{equation}
Detailed forward and inverse surrogate diagnostics, including spatial error profiles and parity
comparisons, are presented and discussed in results section along with Figs.~\ref{fig:appendixC_deeponet} and~\ref{fig:R9}.

\section{Prior Construction for Nozzle Geometry Parameterizations}
\label{app:priors}

This appendix describes the prior distributions imposed on the nozzle area distribution \(A(x)\) for the three geometric parameterizations used in this work: discrete piecewise-linear, seventh-degree polynomial, and cubic B-spline representations.  
All priors enforce exact endpoint value and slope constraints and are designed to generate physically plausible nozzle geometries while enabling a controlled comparison of locality, smoothness, and global coupling effects.  
Fig.~\ref{fig:prior_comparison} summarizes the parameter-space priors and their induced admissible envelopes in physical space. \BT{The parameters appearing in the prior constructions correspond to the geometry parameter vector $\boldsymbol{\theta}$ introduced in Section~\ref{sec:geometry_parameterization}, whose components for each representation are defined explicitly in Appendix~\ref{app:geom_param}.}

\subsection{Discrete Piecewise-Linear Prior}
\label{app:prior_discrete}

In the discrete piecewise-linear representation, the nozzle area is specified at a set of \(K=20\) uniformly spaced axial locations \(\{x_j\}_{j=0}^{K-1}\).  
A baseline geometry \(A_{\mathrm{base}}(x)\) is first defined by linear interpolation between the inlet and outlet areas,
\begin{equation}
A_{\mathrm{base}}(x) = A_0 + \frac{A_L - A_0}{L}\,x ,
\end{equation}
and the corresponding baseline control values are
\begin{equation}
A_j^{\mathrm{base}} = A_{\mathrm{base}}(x_j), \qquad j=0,\ldots,K-1 .
\end{equation}

The endpoint control points are fixed exactly,
\begin{equation}
A_0 = A(0), \qquad A_{K-1} = A(L),
\end{equation}
while the interior control values are treated as independent random variables with uniform bounds,
\begin{equation}
A_j \sim \mathcal{U}\!\left(0.7\,A_j^{\mathrm{base}},\;1.3\,A_j^{\mathrm{base}}\right),
\qquad 1 \le j \le K-2 .
\label{eq:prior_discrete}
\end{equation}

This construction allows each interior control point to vary by \(\pm 30\%\) about the baseline geometry, providing strictly local geometric flexibility.  
However, because the representation is only \(C^0\)-continuous, this prior permits sharp geometric variations, resulting in a relatively wide admissible envelope for \(A(x)\), as shown in Fig.~\ref{fig:prior_comparison}(a) and (b).

\subsection{Seventh-Degree Polynomial Prior}
\label{app:prior_poly}

In the polynomial representation, the area of the nozzle is expressed as a seventh-degree polynomial in the coordinate $x$,
\begin{equation}
A(x) = \sum_{k=0}^{7} b_k x^k .
\end{equation}

Endpoint value and slope constraints uniquely determine the coefficients \(b_0, b_1, b_6,\) and \(b_7\) (see Appendix~\ref{app:geom_param}), leaving the coefficient vector
\[
\boldsymbol{b}_{\mathrm{free}} = (b_2,b_3,b_4,b_5)
\]
as the free parameters.  
A baseline geometry \(A_{\mathrm{base}}(x)\) is implicitly defined by setting each free coefficient to the midpoint of its admissible range.

Independent uniform priors with coefficient-specific bounds are assigned,
\begin{equation}
b_2 \sim \mathcal{U}(-2,\,2), \quad
b_3 \sim \mathcal{U}(-3,\,3), \quad
b_4 \sim \mathcal{U}(-4,\,4), \quad
b_5 \sim \mathcal{U}(-8,\,8).
\label{eq:prior_poly}
\end{equation}

Because polynomial basis functions have global support, perturbations to any coefficient modify the geometry over the entire axial domain.  
Consequently, although this representation enforces global smoothness, it introduces strong nonlocal coupling, leading to wide priors as illustrated in Fig.~\ref{fig:prior_comparison}(c) and (d).

\subsection{Cubic B-Spline Prior}
\label{app:prior_bspline}

The cubic B-spline representation combines smoothness with localized geometric control.  
The nozzle area is expressed as
\begin{equation}
A(x) = \sum_{i=0}^{K-1} N_{i,3}(x)\,c_i ,
\end{equation}
where \(N_{i,3}\) are cubic B-spline basis functions defined on a clamped knot sequence.  
Endpoint interpolation enforces
\begin{equation}
A(0)=c_0=A_0, \qquad A(L)=c_{K-1}=A_L ,
\end{equation}
and the endpoint slopes uniquely determine the adjacent coefficients \(c_1\) and \(c_{K-2}\) (Appendix~\ref{app:geom_param}).

A smooth baseline geometry \(A_{\mathrm{base}}(x)\) is constructed by interpolating between the inlet and outlet areas while accounting for the expected contraction--expansion shape of the nozzle.  
The corresponding baseline control coefficients \(c_i^{\mathrm{base}}\) are obtained by evaluating this reference geometry at the Greville abscissae, thereby encoding both endpoint information and the global geometric trend.

Only the interior coefficients are treated as random variables,
\begin{equation}
c_i = c_i^{\mathrm{base}} + \delta_i,
\qquad
\delta_i \sim \mathcal{U}\!\left(-0.3\,c_i^{\mathrm{base}},\,0.3\,c_i^{\mathrm{base}}\right),
\qquad 2 \le i \le K-3 .
\label{eq:prior_bspline}
\end{equation}

\begin{figure}[H]
  \centering
  \includegraphics[width=\linewidth]{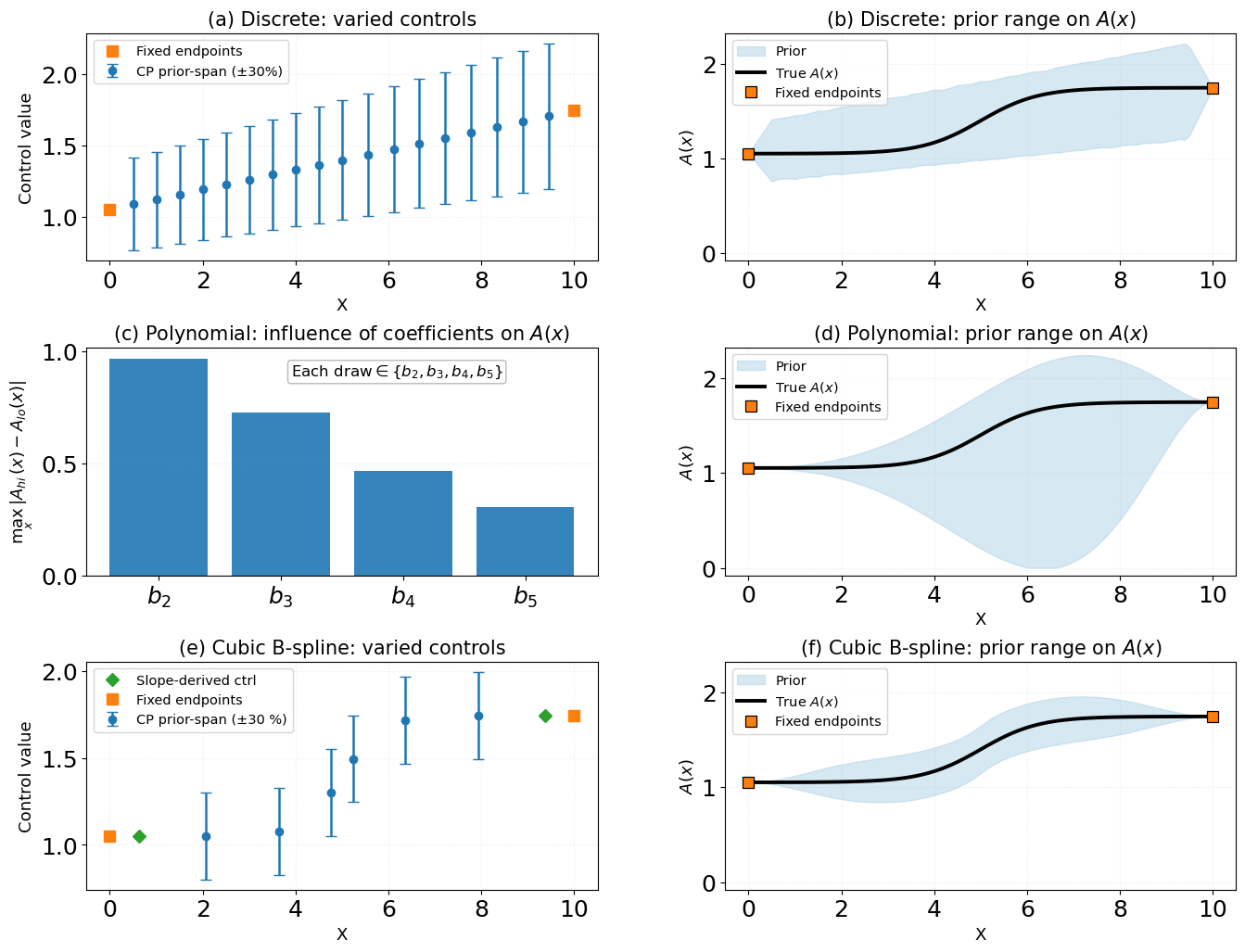}
  \caption{
  Comparison of geometry priors.
  (a), (c), and (e) are parameter-space priors for the discrete piecewise-linear, polynomial, and cubic B-spline representations.
  (b), (d), and (f) are corresponding admissible envelopes for the nozzle area \(A(x)\), with the reference geometry shown in black.
  }
  \label{fig:prior_comparison}
\end{figure}

This \(\pm 30\%\) variation allows the prior to encompass a wide range of physically admissible nozzle shapes required for inverse reconstruction, while preserving \(C^2\) continuity and strong locality.  
As shown in Fig.~\ref{fig:prior_comparison}(e) and (f), the induced envelope remains well controlled compared to the polynomial case.

\subsection{Summary}

The discrete prior provides maximal locality without smoothness; the polynomial prior enforces global smoothness with strong nonlocal coupling; and the cubic B-spline prior achieves smoothness and locality simultaneously through compactly supported basis functions.  
These structural differences directly influence the admissible geometry space explored by the inverse problem and strongly affect numerical conditioning, uncertainty propagation, and surrogate-model training.

\FloatBarrier

\end{appendices}

\bibliographystyle{unsrtnat}
\bibliography{sn-bibliography}

\end{document}